\documentclass[twocolumn,pre,amssymb]{revtex4}
\usepackage{times}
\usepackage{subfigure}
\usepackage[sort&compress]{natbib}
\usepackage{epsfig}
\usepackage{graphicx}
\usepackage{amsmath}

\newcommand{\figref}[1]{Figure~\ref{#1}}

\newcommand\den{\phi}

\def\bar#1{\overline{#1}}

\def \amon{i}
\def \bmon{j}
\def \cmon{k}
\def \amol{a}
\def \abas{\alpha}
\def \bbas{\beta}
\def \sid{\hat{S}}
\def \sinc{\tilde{S}}
\def \srpa{S}
\def \perom{\tilde{\omega}}
\def \perden{\tilde{\den}}
\def \perxi{\tilde{\xi}}
\def \perq{\tilde{q}}

\def \domvec{\boldsymbol{\delta\tilde{\omega}}}
\def \drhovec{\boldsymbol{\delta\tilde{\phi}}}
\def \dxivec{\boldsymbol{\delta\tilde{\xi}}}

\def \kv{{\bf k}}
\def \rv{{\bf r}}
\def \Gv{{\bf G}}
\def \smat{{\bf S}}
\def \sidmat{\hat{{\bf S}}}
\def \sincmat{\tilde{{\bf S}}}

\graphicspath{{figs/}}

\begin{document}

\title{Linear Response and Stability of Ordered Phases of Block Copolymer Melts}

\author{ Amit Ranjan, Jian Qin, and David C. Morse }
  
\affiliation{
Department of Chemical Engineering and Materials Science, 
University of Minnesota, 
421 Washington Ave. S.E., Minneapolis, MN 55455 }
\date{\today}

\begin{abstract}
An efficient pseudo-spectral numerical method is introduced for calculating
a self-consistent field (SCF) approximation for the linear susceptibility 
of ordered phases in block copolymer melts (sometimes referred to as the 
random phase approximation). Our method is significantly more efficient than 
that used in the first calculations of this quantity by Shi, Laradji and 
coworkers, allowing for the study of more strongly segregated structures. 
We have re-examined the stability of several phases of diblock copolymer 
melts, and find that some conclusions of Laradji {\it et al.} regarding 
the stability of the Gyroid phase were the result of insufficient spatial 
resolution. We find that an epitaxial ($\kv=0$) instability of the Gyroid 
phase with respect to the hexagonal phase that was considered previously 
by Matsen competes extremely closely with an instability that occurs at 
a nonzero crystal wavevector $\kv$.
\end{abstract}

\date{\today}
\maketitle

\section{Introduction}
\label{sec:intro}
The local stability of periodic structures, such as those formed by 
block copolymer melts, may be characterized by a linear response function 
that describes the nonlocal response of the monomer concentrations to 
a hypothetical external chemical potential field.  This response 
function is closely related to the correlation function that is probed 
by small angle x-ray and neutron scattering.  The use of a self-consistent 
field (SCF) approximation to calculate the linear susceptibility of 
a homogeneous polymer blend or disordered diblock copolymer melt is 
often referred as a ``random phase approximation" (RPA) for the 
correlation function.\citep{dgbook} This usage has become entrenched 
in the polymer literature, despite its somewhat obscure origin 
\cite{pines1,pines2,pinesnozieres} as one of several names for the use 
of a time-dependent SCF theory ({\it i.e.}, time-dependent Hartree 
theory) to calculate the dynamic linear response of a free electron 
gas \citep{ashmer,ziman_book}.

In this paper, we present an efficient numerical method to calculate 
the SCF linear susceptibility of ordered phases of block copolymers. 
The SCF susceptibility of the disordered phase of a diblock copolymer 
melt was first calculated by Leibler \cite{leibler}. Leibler
used this both to describe diffuse scattering from the disordered phase, 
and as one building block in his theory of weak microphase segregation. 
Shi, Laradji and coworkers \citep{shi_macro1,shi_prl,shi_macro2} were 
the first to calculate the SCF susceptibility of ordered phases of 
diblock copolymer melts, which they used to examine the limits of 
local stability of various ordered structures. 

The calculation of the full linear response for an ordered 
structure is a numerically intensive task. Laradji {\it et al.}
\citep{shi_macro1,shi_prl,shi_macro2} used a spectral method to
complete this calculation that was closely analogous to the algorithm 
used by Matsen and Schick \cite{matsen_schick} to study equilibrium 
structures. When applied to three dimensionally periodic structures 
such as the BCC and Gyroid phases, this algorithm was able to 
accurately describe only very weakly segregated structures. This 
limitation was a result of a rapid increase in computational cost
with increases in the number $M$ of spatial degrees of freedom 
required to resolve the structure: The computational cost of a 
spectral calculation of the response to a single perturbation 
({\it e.g.}, the response to a single plane wave) is of 
${\cal O}(M^{3})$, while the cost of a calculation of the full 
RPA response function ({\it i.e.}, the response to an arbitrary 
small perturbation) is ${\cal O}(M^{4})$.

The earlier use of a spectral method by Matsen and Schick\cite{matsen_schick}
to calculate the equilibrium phase diagram for diblock copolymer melts relied
heavily upon the use of space group symmetry to decrease the number of basis
function needed to describe a structure. Matsen and Schick introduced the use
of basis functions with the space group symmetry of each structure of interest.
Use of these symmetry-adapted basis functions functions reduces the number of
degrees of freedom $M$ needed to obtain a given spatial resolution by a factor
roughly equal to the number of point group symmetries in the relevant space
group. For the BCC and Gyroid cubic phases, this reduces $M$ by almost a factor
of 48, and thus reduced the cost of solution a single iteration of the SCF
equations by a factor of almost $(48)^{3} \approx 10^{5}$.  

The calculation of the full linear susceptibility, however, requires 
the calculation of the response to arbitrary infinitesimal perturbations, 
which generally do not preserve the space group symmetry of the crystal. 
In general, it thus requires the use of either a plane wave basis or 
a spatial discretization that does not impose any symmetry upon the 
perturbation.  Primarily as a result of this loss of the advantages
of symmetry, Laradji {\it et al.}\cite{shi_prl,shi_macro2} were able
to obtain results for the Gyroid phase only for $\chi N \leq 12$.
Matsen has been able to carry out linear response calculations for 
the BCC and Gyroid phases at significantly larger values of $\chi N$ 
by considering only the response to perturbations that preserve the 
periodicity of the cubic lattice, and that preserve a subgroup of the 
space group of the unperturbed structure \cite{cg_matsen,cs_matsen}. 
Matsen's method and results are discussed in more detail in Sec. 
\ref{sec:diblock}.

More recent work on numerical methods for solving the equilibrium 
SCFT by Rasmussen and Kalosakas \cite{kalosakas}, and by Fredrickson 
and co-workers \cite{sides_fred3,glenn_gyr,glennbook} has made use 
of pseudo-spectral methods for which the cost of a single iteration 
of the SCFT equations scales as ${\cal O}(M \ln M )$, rather than 
${\cal O}(M^{3})$. Here, we present a derivation of a perturbation 
theory for the linear response to an arbitrary disturbance in a form 
that we can evaluate numerically using a pseudo-spectral algorithm. 

The remainder of the paper is organized as follows: 
In Section~\ref{sec:Bloch}, we discuss the basic formalism of 
the SCF linear response theory for perturbations of a periodic 
microstructure, and review the consequences of Bloch's theorem. 
In Section~\ref{sec:perturbation}, we present a perturbation 
theory for the underlying modified diffusion equation, which is 
used to calculate the linear susceptibility of a reference system 
of non-interacting polymers in a periodic potential.  
In Section~\ref{sec:matrixform}, we discuss the use of linearized
SCFT to calculate the corresponding susceptibility of an 
incompressible system of interacting chains.  In Section~\ref{eff}, 
we discuss the implementation and efficiency of our algorithm.
Section~\ref{sec:diblock} presents selected results regarding 
the stability of the HEX, BCC, and (particularly) the Gyroid 
phases of diblock copolymer melts.

\section{Response of a periodic structure}
\label{sec:Bloch}
In self-consistent field theory, we calculate the average 
concentration $\rho_{\amon}(\rv)$ for monomers of type $\amon$ 
by considering a hypothetical system of non-interacting polymers 
in which monomers of type $\amon$ are subjected to a 
self-consistently determined field $\omega_{\amon}(\rv)$. 
We consider incompressible systems in which each monomer 
occupies a volume $v$, and define a volume fraction field 
$\phi_{\bmon}(\rv) \equiv v \rho_{\bmon}(\rv)$.

In what follows, we consider the response of an unperturbed state in 
which $\omega_{\amon}(\rv)$ satisfies the self-consistent field 
(SCF) equation
\begin{equation}
\label{internalfield}
    \omega_\amon(\rv) = \chi_{\amon\bmon}\den_{\bmon}(\rv) +
    \xi(\rv) \quad.
\end{equation}
Here, $\chi_{\amon\bmon}$ is a Flory-Huggins parameter for binary 
interactions between monomers of types $\amon$ and $\bmon$, and 
$\xi(\rv)$ is a Lagrange multiplier field that must be chosen so 
as to satisfy an incompressibility constraint. 
To calculate the SCF linear susceptibility, we consider the 
perturbation caused by an additional infinitesimal external potential 
$\delta\omega^{\rm ext}_{\amon}(\rv)$. The resulting deviation 
$\delta\omega_{\amon}$ must satisfy
\begin{equation} 
    \delta \omega_\amon(\rv) =
    \chi_{\amon\bmon} \delta\den_{\bmon}(\rv) 
   + \delta\xi(\rv)
   + \delta\omega^{\rm ext}_\amon(\rv)
\end{equation}
where $\delta\den_{\amon}(\rv)$ is the corresponding deviation
in the monomer concentration, and $\delta\xi(\rv)$ is the
deviation in $\xi(\rv)$ required to satisfy the incompressibility
constraint
\begin{equation}
   \label{incomp}
   \sum_{\amon}\delta\den_{\amon}(\rv) = 0 
\end{equation}

The deviation $\delta\den_{\amon}(\rv)$ may be described by 
either of two related linear response functions: It may be expressed
either as an integral
\begin{equation}
   \label{sdef_{\amol}deal}
    \delta\den_{\amon}(\rv) =  
    - \int \sid_{\amon\bmon}(\rv,\rv')
    \delta\omega_{\bmon}({\bf r'}) d\rv' 
\end{equation}
in which $\sid_{\amon\bmon}(\rv,\rv')$ is nonlocal susceptibility of 
an inhomogeneous gas of non-interacting polymers to a change in the 
total self-consistent field $\omega$, or by a corresponding relation 
\begin{equation}
   \label{sdef_rpa}
   \delta\den_{\amon}(\rv) =  
   - \int \srpa_{\amon\bmon}(\rv,\rv')
   \delta\omega^{\rm ext}_{\bmon}({\bf r'}) d\rv' 
\end{equation}
in which $\srpa_{\amon\bmon}(\rv,\rv')$ is the SCF susceptibility of 
the interacting liquid of interest to the external potential 
$\omega^{\rm ext}$.

It is convenient to introduce a Fourier representation of the problem.
As an example, we consider the ideal gas response function $\sid$ in
what follows, but identical arguments apply to $\srpa$.  The linear 
response to a perturbation
\begin{equation}
   \delta\omega_{\amon}(\rv) 
    = \sum_{\kv}e^{i\kv\cdot\rv} \delta\omega_{\amon}(\kv)
\end{equation}
may be expressed in Fourier space, for a system of finite volume, as 
a sum
\begin{equation}
    \delta\den_{\amon}(\kv) = \sum_{\kv'}
    \sid_{\amon\bmon}(\kv,\kv')
    \delta\omega_{\bmon}({\bf \kv'}) 
\end{equation}
in which
\begin{equation}
    \sid_{\amon\bmon}(\kv,\kv') \equiv
    \frac{1}{V}\int {\bf dr} \int {\bf dr}'
    \sid_{\amon\bmon}(\rv,\rv')
    e^{-i\kv\cdot\rv + i\kv'\cdot\rv'}
    \label{sidk_def}
\end{equation}
where $V$ is the volume of a system containing many unit cells of 
the original structure, with Born-von Karmann boundary conditions.  

We are interested here in the response of an unperturbed structure 
that is invariant under translations $\rv \rightarrow \rv+{\bf R}$, 
where ${\bf R}$ is any vector in the Bravais lattice of the unperturbed 
crystal. As a result, we expect any linear response function to exhibit 
the symmetry
\begin{equation}
    \sid_{\amon\bmon}(\rv,\rv') \equiv
    \sid_{\amon\bmon}(\rv + {\bf R}, \rv' + {\bf R}) 
\end{equation}
for any lattice vector ${\bf R}$. 
By Fourier transforming both sides of this equality, using definition
(\ref{sidk_def}) for the transform, we find that 
$\sid_{\amon\bmon}(\kv,\kv')$ can be nonzero only for values of $\kv$ 
and $\kv'$ for which $e^{i(\kv - \kv')\cdot{\bf R}} = 1$ for any 
lattice vector ${\bf R}$.  The reciprocal lattice is the set of all
wavevectors ${\bf G}$ such that $e^{i{\bf G}\cdot{\bf R}} =1$ for 
any Bravais lattice vector ${\bf R}$.  This implies that 
$\sid_{\amon\bmon}(\kv,\kv')$ can be nonzero only for values of $\kv$ 
and $\kv'$ for which $\kv - \kv' = {\bf G}$ for some reciprocal lattice 
vector ${\bf G}$.  This is the content of Bloch's theorem, as applied 
to a linear response function.  It is thus convenient to represent the 
nonzero elements of $\sid$ by a matrix
\begin{equation}
   \sid_{\amon\bmon}({\bf G},{\bf G}';\kv) \equiv
   \sid_{\amon\bmon}({\bf G}+\kv,{\bf G}'+\kv)
\end{equation}
where $\kv$ is a crystal wavevector in the first Brillouin zone. 
A similar notation will be used for the SCF response function $\srpa$.  

Consider the response to perturbation that has the form of a Bloch 
function, 
\begin{equation}
   \delta\omega_{\amon}(\rv) = 
   \delta{\perom}_{\amon}(\rv)
   e^{i\kv\cdot\rv}
\end{equation}
in which $\kv$ is a crystal wavevector in the first
Brillouin zone, and $\delta{\perom}_{\amon}(\rv)$ is a 
periodic function with the periodicity of the unperturbed 
lattice,
\begin{equation}
    \delta{\perom}_{\amon}(\rv) = 
    \sum_{{\bf G}}
    \delta{\perom}_{\amon}({\bf G}) e^{i{\bf G}\cdot\rv}
\end{equation}
in which $\sum_{{\bf G}}$ denotes a sum over reciprocal lattice 
vectors. Bloch's theorem guarantees that the resulting density 
perturbation will assume the same form
\begin{equation}
   \delta{\den}_{\amon}(\rv) = 
   \delta{\perden}_{\amon}(\rv)e^{i\kv\cdot\rv}
\end{equation}
where $\delta{\perden}(\rv)$ is also a periodic function.
The relationship between the Fourier components 
$\delta{\perom}_{\amon}({\bf G})$ and
$\delta{\perden}_{\amon}({\bf G})$ may be expressed as a matrix 
product
\begin{equation}
    \delta{\perden}_{\amon}({\bf G}) = 
    - \sum_{{\bf G}'\bmon}
    \sid_{\amon\bmon}({\bf G},{\bf G}';\kv)
    \delta\perom_{\bmon}({\bf G}')
    \label{sdef1}
\end{equation}
with a  matrix $\sid_{\amon\bmon}({\bf G},{\bf G}';\kv)$
whose elements depend parametrically upon $\kv$. It also 
follows from the block-diagonal form of $\sid(\kv,\kv')$ 
that the eigenmodes of $\sid$ and $\srpa$ must be Bloch 
functions. As emphasized by Shi \cite{shi_jph}, these
conclusions about the consequences of periodicity are quite 
general, and are independent of the self-consistent field 
approximation. 

The SCF response function is related to the SCFT free energy
functional $F[\den]$ by the standard identity
\begin{equation}
   \srpa_{\amon\bmon}^{-1}({\bf G},{\bf G}';\kv)
   = \frac{\delta^{2} F[\den]}
     {\delta\den_{\amon}(\kv+{\bf G})
      \delta\den_{\bmon}(-\kv-{\bf G}')}
\end{equation}
Here, the inverse of $\srpa$ is defined in reciprocal space
by requiring
\begin{equation}
  \sum_{\bmon,{\bf G'}}
  \srpa_{\amon\bmon}^{-1}({\bf G},{\bf G}';\kv)
  \srpa_{\bmon\cmon}({\bf G}',{\bf G}'';\kv) =
  \delta_{\amon\cmon}\delta_{{\bf G},{\bf G}''}
\end{equation}
The condition for local stability of a periodic structure is thus 
that the matrix $\srpa_{\amon\bmon}^{-1}({\bf G},{\bf G}';\kv)$ 
be positive definite for every $\kv$ in the first Brillouin 
zone.  The onset of instability occurs when one of the eigenvalues 
of $\srpa_{\amon\bmon}^{-1}({\bf G},{\bf G}';\kv)$ passes 
through zero at some $\kv$.
 
\section{Ideal Gas Response}
\label{sec:perturbation}
In SCFT, the monomer concentration fields are obtained by 
calculating the concentrations in a reference system of non-interacting 
chains in which monomers of type $\amon$ are subjected to a field 
$\omega_{\amon}(\rv)$. This reference system is treated by considering 
a pair of constrained partition functions $q_{\amol}(\rv,s)$ and 
$q^{\dagger}_{\amol}(\rv,s)$ for chains 
of species $\amol$, which satisfy the modified diffusion equations
\begin{eqnarray}
  \frac{\partial q_{\amol}}{\partial s}  & = & - H q_{\amol}
  \nonumber \\
  \label{diffeq}
  \frac{\partial q_{\amol}^{\dagger}}{\partial s} & = & + H q_{\amol}^{\dagger}
  \label{diffeq-dagger}
\end{eqnarray}
in which 
\begin{equation}
    \label{hamil}
    H \equiv -\frac{b_{\amon}^2}{6}\nabla^2+\omega_{\amon}(\rv)
\end{equation}
These quantities satisfy initial conditions $q(\rv,s=0)=1$ and 
$q^{\dagger}(\rv,s=N_{\amol})=1$, respectively, where $N_{\amol}$ 
is the length of chains of type $\amol$.  Here, $b_{\amon}$ 
and $\omega_{\amon}$ are a statistical segment length and a chemical 
potential field for monomers of the type $\amon$ found at point 
$s$ along the chains of type $a$.  The volume fraction of monomers 
of type $\amon$ on chains of type $a$ is given by an integral
\begin{equation}
  \label{rhochain}
  \den_{\amol\amon}(\rv) =
  \frac{\overline{\phi}_{\amol}}{N_{\amol} Q_{\amol}}
  \int \! ds \; q_{\amol}(\rv,s) q^{\dagger}_{\amol}(\rv,s)
  \quad,
\end{equation}
where the integral with respect to $s$ is taken only over those 
blocks that contain monomers of type $\amon$. Here, 
$\bar{\phi}_{\amol}$ is the volume fraction of chains of 
type $\amol$, and 
\begin{equation}
  \label{pf}
  Q_{\amol} \equiv \frac{1}{V}\int\! d\rv \; q_{\amol}(\rv,N_{\amol})
  \quad.
\end{equation}  
The chemical potential $\mu_{\amol}$ for species $i$ and $Q_{\amol}$ are 
connected (for a particular choice of standard state) by a 
relation
\begin{equation}
  \overline{\phi}_{\amol} = Q_{\amol} e^{\mu_{\amol}/kT} 
  \quad. \label{phi_mu}
\end{equation}
The SCFT can be applied in either the canonical 
or grand-canonical ensemble by simply regarding either 
$\overline{\phi}_{\amol}$ or $\mu_{\amol}$ as a specified input
parameter, respectively.
 
Here, we consider a perturbation theory for the variation in 
$q_{\amol}$ that results from a variation in $\omega_{\amon}$. 
The chemical potential field can be written as a sum 
\begin{equation}
  \label{delomega}
  \omega_{\amon}(\rv,s) = \omega_{\amon}^{(0)}(\rv,s) +
  \delta\omega_{\amon}(\rv,s)
\end{equation}
of an unperturbed part $\omega^{(0)}_{\amon}(\rv)$ and a 
small perturbation $\delta\omega_{\amon}(\rv)$.  Similarly 
$q_{\amol}(\rv,s)$ can be expressed as a sum
\begin{equation}
  \label{delq}
  q_{\amol}(\rv,s) = q_{\amol}^{(0)}(\rv,s) + \delta q_{\amol}(\rv,s)
\end{equation}
where $q_{\amol}^{(0)}(\rv,s)$ is the solution to the SCF equations
for the unperturbed periodic structure.  Substituting equations
\ref{delomega} and \ref{delq} in \ref{diffeq} yields the perturbation 
equation:
\begin{equation}
    \label{perteq}
    \left ( \frac{\partial}{\partial s} + H^{(0)} \right) 
    \delta q_{\amol}(\rv,s) = 
    -\delta\omega_{\amon}(\rv)q_{\amol}^{(0)}(\rv,s)
  \end{equation}
where $H^{(0)}$ is the unperturbed ``Hamiltonian". 
This must be be solved subject to an initial condition
$\delta q_{\amol}(\rv,0) = 0$. The unperturbed fields 
$\omega^{(0)}(\rv)$ and $q_{\amol}^{(0)}(\rv,s)$ have the 
periodicity and space group symmetry of the unperturbed crystal. 

To take advantage of Bloch's theorem, we consider a perturbation 
of the form of a Bloch function
$\delta\omega_{\amon}(\rv)= e^{i\kv\cdot\rv}
\delta\perom_{\amon}(\rv)$, 
which will produce a corresponding deviation
\begin{equation}
   \delta q_{\amol}(\rv,s) = 
   e^{i\kv\cdot\rv}\delta\perq_{\amol}(\rv,s)
\end{equation}
Here, $\kv$ is a wavevector in the first Brillouin zone, 
and $\delta\perom_{\amon}$ and $\delta\perq_{\amol}$ are periodic 
functions.  Substituting these expressions into the modified 
diffusion equation and keeping terms that are linear in the
perturbation yields an inhomogeneous PDE:
\begin{equation}
 \label{per_de}
 \left ( \frac{\partial}{\partial s}
    + H_{\kv} \right ) \delta \perq_{\amol}(\rv,s)
    = -\delta\perom_{\amon}(\rv)q_{\amol}^{(0)}(\rv,s) 
\end{equation}
in which
\begin{equation}
   H_{\kv} \equiv 
  -\frac{b^2_{\amon}}{6}(\nabla + i\kv)^2 
  + \omega^{(0)}_{\amon}(\rv)
   \quad.
\end{equation}
A closely analogous PDE may be obtained for 
$\delta \perq^{\dagger}_{\amol}(\rv,s)$.
The resulting deviations must satisfy boundary conditions 
$\delta \perq_{\amol}(\rv,0)=0$ and
$\delta \perq^{\dagger}_{\amol}(\rv,N_{a})=0$ where $N_{a}$
is the number of monomers in a chain of type $a$. 
To calculate the ideal gas linear response, we numerically solve 
this pair of PDEs using a pseudo-spectral algorithm that is
presented in the appendix. 

The perturbation in the periodic part of the monomer concentration 
field may be expressed, in grand-canonical ensemble, as an integral
\begin{equation}
  \label{delrho}
  \delta{\perden}_{\amol\amon}(\rv) =
        \frac{\overline{\phi_{\amol}}}{Q_{\amol}}
        \int \frac{ds}{N_{\amol}} \left[  
        \delta{\perq_{\amol}}(\rv,s) 
        q_{\amol}^{\dagger}(\rv,s)
      + q_{\amol}(\rv,s)\delta{\perq_{\amol}}^{\dagger}(\rv,s)
        \right] 
\end{equation}
Here, $\bar{\phi}_{\amol}$, $Q_{\amol}$, $q_{\amol}(\rv,s)$ and 
$q_{\amol}^{\dagger}(\rv,s)$ all represent values evaluated in 
the unperturbed state. The integral with respect to $s$ in the above 
must be taken only over the block or blocks that contain monomers of 
type $\amon$. 

The expression for $\delta{\perden}_{\amol\amon}(\rv)$ in 
canonical ensemble is the same as that obtained for grand canonical 
ensemble for any crystal wavevector $\kv$ {\it except} $\kv=0$. 
It may be shown that only perturbations with $\kv=0$ ({\it i.e.}, 
perturbations with the same periodicity as the unperturbed crystal) 
can induce changes in $Q_{\amol}$ to linear order in the strength 
of the applied potential. In grand-canonical ensemble, any change 
$\delta Q_{\amol}$ in $Q_{\amol}$ will cause a change $\delta\bar{\phi}_{\amol}
= \delta Q_{\amol}e^{\mu_{\amol}/kT}$ in the molecular volume 
fraction $\bar{\phi}_{\amol}$ obtained from Eq. (\ref{phi_mu}), 
but the prefactor to $e^{\mu_{\amol}} = \bar{\phi}_{\amol}/Q_{\amol}$ in 
Eq. (\ref{rhochain}) remains constant.  In canonical ensemble, where 
$\bar{\phi}_{\amol}$ is regarded as an fixed input parameter, a change 
in $Q_{\amol}$ instead induces a change in the denominator of Eq. 
(\ref{rhochain}) for ${\perden}_{\amol\amon}(\rv)$. This yields 
a slightly modified expression
\begin{equation}
  \delta{\perden}_{\amol\amon}(\rv) =
  {\rm GCE} - \den_{\amol\amon}^{(0)}(\rv) \frac{\delta Q_{\amol}}{Q_{\amol}}
\end{equation}
for perturbations in canonical ensemble at exactly $\kv=0$, in 
which ``GCE" represents the grand-canonical ensemble response given 
by the right hand side of Eq. (\ref{delrho}). It may be shown that 
this expression yields a perturbation in which 
$\int d\rv\;\delta{\perden}_{\amol\amon}(\rv) =0$ for all 
$\amol$ and $\amon$. 

By using the above perturbation theory to calculate the Fourier
components of the perturbation $\delta\perden_{\bmon}(\rv)$ caused 
by a particular plane wave perturbation 
$\delta\perom_{\amon}(\rv) \propto e^{i{\bf G}'\cdot\rv}$, we may 
obtain one row of the matrix $\sid_{\amon\bmon}({\bf G,G'};\kv)$
at a specified value of $\kv$.  The elements of this reciprocal-space 
matrix are generally complex numbers, but may be shown to be real 
when the unperturbed crystal has inversion symmetry.

\section{Self-Consistent Response}
\label{sec:matrixform}
We now discuss how the SCF susceptibility 
$\srpa({\bf G},{\bf G}';\kv)$ can be calculated from the response 
function $\sid({\bf G},{\bf G}',\kv)$ of an ideal gas. 

\subsection{General Analysis}
Consider the response to an external perturbation of the Bloch 
form $\delta\omega_{\amon}^{\rm ext} (\rv) =  
\delta \perom_{\amon}^{\rm ext}(\rv)e^{i\kv\cdot\rv}$. 
Substituting the self-consistency condition into definition 
\ref{sdef1} of the ideal gas response function, and expressing the 
result in Fourier space, yields the linear self-consistency condition
\begin{equation}
     \label{rpa0}
     \delta \perden_{\amon}({\bf G})
     = - \sid_{\amon\bmon}({\bf G},{\bf G}';\kv)
     \delta\perom_{\bmon}({\bf G}') 
\end{equation}
where
\begin{equation}
    \delta\perom_{\bmon}({\bf G}')  = 
    \delta\perom^{\rm ext}_{\bmon}({\bf G}') + 
    \chi_{\bmon\cmon}\delta\perden_{\cmon}({\bf G}') +
    \epsilon^{+}_{\bmon} \delta\perxi({\bf G}') 
\end{equation}
Summation over repeated reciprocal wavevectors and monomer 
type indices is implicit.  Here we have introduced the notation 
$\epsilon^{+}_{\bmon}$ for a vector for which $\epsilon^{+}_{\bmon} =1$ 
for all $\bmon$, and $\delta\perxi({\bf G}')$ for a Fourier component 
of the periodic part $\delta\perxi(\rv)$ of a deviation
\begin{equation}
  \delta \xi(\rv) = \delta\perxi(\rv) e^{i\kv\cdot\rv}
\end{equation}
Eq. (\ref{rpa0}) can be expressed more compactly, in a matrix notation, 
as
\begin{eqnarray}
     \label{rpa1}
     \drhovec_{\amon} & = & 
     - \sidmat_{\amon\bmon}
     {\domvec}_{\bmon} \nonumber \\
     & = & - \sidmat_{\amon\bmon} 
     \left[ \domvec^{\rm ext}_{\bmon} + 
     \chi_{\bmon\cmon}\drhovec_{\cmon} +
     \epsilon^+_{\bmon}\dxivec\right]
\end{eqnarray}
Here and hereafter, we use boldfaced Greek letters with a single latin
monomer index $i,j,k,\ldots$ to represent column vectors in the space 
of reciprocal lattice vectors (or periodic functions of $\rv$), so 
that ${\drhovec}_{\amon} \equiv \delta{\perden}_{\amon}({\bf G})$ and
${\domvec}_{\bmon} \equiv {\delta\perom}_{\bmon}({\bf G'})$, and 
boldfaced capital Roman letter with two monomer type indices to 
represent matrices in this space, so that 
$\sidmat_{\amon\bmon} \equiv \sid_{\amon\bmon}({\bf G},{\bf G'};\kv)$. 
In this notation, matrix-vector and matrix-matrix multiplication is thus 
used to represent summation over repeated reciprocal lattice vector arguments.
When monomer types indices are displayed explicitly, summation over repeated
indices is implied.

Imposing the incompressibility constraint
\begin{equation}
   0 = \sum_{\amon} \delta \perden_{\amon}({\bf G}) 
     = \epsilon^{+}_{\amon}\delta \perden_{\amon}({\bf G})
   \label{incompressible_G}
\end{equation}
yields a condition
\begin{equation}
     \label{inc_mat}
     0 = 
     \epsilon_{\amon}^{+} \sidmat_{\amon\bmon}
     \left[ \domvec^{\rm ext}_{\bmon} + 
     \chi_{\bmon\cmon}\drhovec_{\cmon} +
     \epsilon^+_{\bmon}\dxivec\right]
\end{equation}
Solving Eq. (\ref{inc_mat}) for $\dxivec$ yields
\begin{equation}
    \label{dxivec}
    \dxivec = - \sidmat^{-1}_{++} 
    \sidmat_{+\bmon} 
    \left[ \domvec^{\rm ext}_{\bmon} +
    \chi_{\bmon\cmon}\drhovec_{\cmon}\right]
    \quad,
\end{equation}
Here, we have introduced the quantities
\begin{eqnarray}
  \sid_{\amon+}({\bf G},{\bf G}';\kv) & \equiv & 
  \sid_{\amon\bmon}({\bf G},{\bf G}';\kv)
  \epsilon_{\bmon}^{+}
  \nonumber \\
  \sid_{+\bmon}({\bf G},{\bf G}';\kv)  & \equiv & 
  \epsilon_{\amon}^{+}
  \sid_{\amon\bmon}({\bf G},{\bf G}';\kv)
  \\
  \sid_{++}({\bf G},{\bf G}';\kv)      & \equiv & 
  \epsilon_{\amon}^{+}
  \sid_{\amon\bmon}({\bf G},{\bf G}';\kv)
  \epsilon_{\bmon}^{+} 
  \nonumber
  \quad. \label{sid_pm}
\end{eqnarray}
These are represented in matrix notation by $\sidmat_{\amon+}$, 
$\sidmat_{+\bmon}$, and $\sidmat_{++}$, respectively. 
Substituting Equation (\ref{dxivec}) for $\dxivec$ back into 
Equation (\ref{rpa1}) yields
\begin{equation}
    \drhovec_{\amon} = - \sincmat_{\amon\bmon}
    \left[ \domvec^{\rm ext}_{\bmon} + 
           \chi_{\bmon\cmon}\drhovec_{\cmon}
    \right ]
    \label{drho_wo_dxi}
\end{equation}
where  
\begin{equation}
    \sincmat_{\amon\bmon} = \sidmat_{\amon\bmon} -
    \sidmat_{\amon+}
    \sidmat^{-1}_{++}
    \sidmat_{+\bmon}
    \label{sincmat_def}
\end{equation}
The quantity $\sincmat_{\amon\bmon}$ is the SCFT response function 
of incompressible system with $\chi_{\amon\bmon} = 0$.  

By solving Eq. (\ref{drho_wo_dxi}), we find that
\begin{equation}
    \drhovec_{\amon} = -\smat_{\amon\bmon}
    \domvec^{\rm ext}_{\bmon} 
\end{equation}
where
\begin{equation}
    \label{rpagen}
    \smat_{\amon\bmon} = 
    \sincmat_{\amon\cmon}
    \left[ {\bf I} - 
    \mbox{${\mathbf \chi}$}\sincmat \right]^{-1}_{\cmon\bmon}
\end{equation}
is the desired SCF response function.
Here ${\bf I}$ denotes the identity in the space of reciprocal 
lattice vectors and monomer type indices, with elements 
$\delta_{\cmon\bmon}\delta_{{\bf G},{\bf G}'}$, 
$\chi\sincmat$ denotes a matrix in this space with elements
$\sum_{\bmon}\chi_{\amon\bmon}\sinc_{\bmon\cmon}({\bf G},{\bf G'})$,
and inversion is defined in this expanded space.
 
In order for the incompressibility condition (\ref{incompressible_G}) to be 
satisfied for the response to an arbitrary infinitesimal perturbation,
the response function $\srpa$ for any incompressible liquid must satisfy
\begin{equation}
    \label{sincompB}
    0 = 
    \srpa_{+\bmon}({\bf G},{\bf G^{\prime}};\kv)
    = 
    \srpa_{\amon +}({\bf G},{\bf G^{\prime}};\kv) 
\end{equation}
for any $\kv$, ${\bf G}$, and ${\bf G'}$. The second equality in 
the above follows from Onsager reciprocity. Eq. \ref{sincompB}
also implies that $\srpa_{++}({\bf G},{\bf G^{\prime}};\kv)=0$. The 
same conditions apply to $\sincmat$, which is a special case of 
$\smat$ for $\chi=0$.  It is straightforward to confirm that Eq. 
(\ref{sincmat_def}) satisfies these conditions for $\sincmat$, and 
that they are preserved by Eq.  (\ref{rpagen}) for $\smat$. 

Eq. (\ref{sincompB}) implies that $\smat$ is singular, if viewed as 
a matrix in the space of reciprocal lattice vectors and monomer types, 
since it implies that
\begin{equation}
   \srpa_{\amon\bmon}({\bf G},{\bf G}') \psi_{\bmon}({\bf G}') = 0
\end{equation}
for any vector of the form
$\psi_{\bmon}({\bf G}') = \epsilon^{+}_{\bmon}\psi({\bf G}')$, for
which the elements are functions of ${\bf G}'$ alone, independent of 
the value of the monomer index $\bmon$. The matrix $\smat_{\amon\bmon}$ 
thus has a null space spanned by the space of all such vectors.
In any numerical calculation for a system of $C$ monomer types in 
which we use a truncated Fourier representation of $M$ reciprocal 
lattice vectors, $\smat$ is a matrix in a $CM$ dimensional space 
that contains an $M$ dimensional null space (or kernel).  Thus, 
though it is tempting for us to rewrite Equation~(\ref{rpagen}) as 
$\smat^{-1} = \sincmat^{-1} - \chi$, this would be meaningless,
because neither $\smat$ nor $\sincmat$ are invertible.  The 
non-null space of the symmetric matrix $\smat$ (i.e., the space 
spanned by all eigenvectors of $\smat$ with non-zero eigenvalues) 
is spanned by all functions for which $\sum_{\amon}\psi_{\amon}({\bf G}) 
= 0$, since this is the condition of orthogonality with any vector 
in the null space. The non-null space is thus the same as the 
space of monomer concentration fluctuations that respect 
incompressibility constraint (\ref{incompressible_G}). 

\subsection{Systems with Two Types of Monomer}
Consider a system containing only two types of monomer, with a 
single interaction parameter $\chi = \chi_{12} = \chi_{21}$. In this
case, it is straightforward to project the problem onto the space of 
physically allowable fluctuations, which respect the incompressibility 
constraint. We define the two component vectors 
$\varepsilon^{\pm}_{\amon}$ = ($1,\pm1$), where 
$\amon=1$ or 2 for the two monomer types, and introduce the following 
transformation for any $2N\times 2N$ dimensional correlation function 
matrix $\srpa_{\amon\bmon}({\bf G},{\bf G}')$:
\begin{equation}
    \label{tranf}
    \srpa_{\mu\nu} ({\bf G},{\bf G'};\kv) =
    \varepsilon^{\mu}_{\amon} 
    \srpa_{\amon\bmon} ({\bf G},{\bf G}';\kv)
    \varepsilon^{\nu}_{\bmon}
    \quad,
\end{equation}
in which $\mu$ and $\nu$ belong to the set $\{+,-\}$.  The quantity
$\sid_{++}({\bf G},{\bf G}')$ defined in Eq. (\ref{sid_pm}) is an 
example of this notation. As already noted, the incompressibility 
constraint requires that $\sincmat_{++} = \sincmat_{+-} = 
\sincmat_{-+} = 0$.  By evaluating the remaining $--$ element of 
Eq.  (\ref{sincmat_def}) for $\sincmat_{\amon\bmon}$, we find that
\begin{equation}
   \label{smm}
   \sincmat_{--} = \sidmat_{--} -
   \sidmat_{-+}\sidmat^{-1}_{++}\sidmat_{+-}
\end{equation}
It is straightforward to show that $\smat_{--}$ and $\sincmat_{--}$ 
are related by
\begin{equation}
    \smat_{--} = \sincmat_{--}
    \left[ {\bf I}-\frac{\chi}{2}\sincmat_{--}\right]^{-1}
    \label{smatmm}
\end{equation}
where $\chi \equiv \chi_{12}$ is the conventional scalar Flory-Huggins 
parameter.  This result is equivalent to Eqs. (17) and (18) of Laradji 
et {\it al.} \cite{shi_macro2}.  The matrix $\sincmat_{--}$ in a 
system described by a truncated basis of $M$ reciprocal lattice vectors 
and $2$ monomer types is an $M\times M$ matrix, which is generally 
non-singular. It therefore {\it is} legitimate to rewrite Eq. 
(\ref{smatmm}) as
\begin{equation}
  \label{smminv}
  \smat^{-1}_{--} = \sincmat^{-1}_{--} - \frac{\chi}{2}{\bf I}
  \quad,
\end{equation}
in close analogy to the RPA equation for an incompressible disordered 
system.

The limits of stability of an ordered structure may be identified by 
examining the eigenvalues of the RPA response function $\smat_{--}$.
At each $\kv$ in the first Brillouin zone, we consider the 
eigenvalue equation
\begin{equation}
   \sum_{{\bf G}'}
   \srpa_{--}({\bf G},{\bf G}';\kv) \psi_{n}({\bf G}';\kv) 
   = \lambda_{n}(\kv) \psi_{n}({\bf G};\kv)
\end{equation}
with eigenvectors $f_{n}({\bf G};\kv)$. The vector 
$\psi_{n}({\bf G};\kv)$ contains the Fourier components of a periodic 
function $\psi_{n}(\rv;\kv)$ that is the periodic part of an eigenvector 
of the Bloch form $e^{i\kv\cdot\rv}\psi_{n}(\rv)$.  It follows from Eq. 
(\ref{smminv}) that $\smat_{--}$ and $\sincmat_{--}$ have the same 
eigenvectors, and that their eigenvalues are related by
\begin{equation}
   \label{evalinv}
   \lambda_{n}^{-1}(\kv) = 
   \tilde{\lambda}_{n}^{-1}(\kv) - \frac{\chi}{2}
\end{equation}
where $\tilde{\lambda}_{n}(\kv)$ is a corresponding eigenvalue 
of the response function matrix $\sincmat_{--}$. The problem of 
calculating the eigenvalues of $\smat$ thus reduces to that of 
calculating and diagonalizing $\sincmat$.

\subsection{Response at $\kv=0$}
When examining limits of stability we will often be particularly
interested in instabilities at $\kv=0$. In the cases of interest,
these correspond to instabilities toward structures that have an 
epitaxial relationship with the original structure, but in which 
some of the reciprocal lattice vectors of the original structure 
are absent in the final structure. In diblock copolymer melts, the 
instabilities of the BCC phase towards hexagonally packed cylinders, 
and of the cylinder phase towards a lamellar phase are found to be 
epitaxial instabilities of this type. The linear response at exactly 
$\kv = 0$, however, has some special features that are important 
to understand when constructing a numerical algorithm.

The response $\srpa_{\amon\bmon}({\bf G},0;0)$ to a perturbation 
at $\kv = {\bf G}' = 0$ corresponds to a response to a spatially
homogeneous shift in monomer chemical potentials. The change in 
potential energy associated with any spatially homogeneous 
perturbation $\delta\omega_{\bmon}^{\rm ext}$ in the external 
fields that couple to monomer concentration depends only upon the
total number of monomers of each type in the system, which can 
change only as a result in the change in the number of molecules 
of each species. Such a homogeneous perturbation is thus equivalent 
to the response to a shift 
$\delta \mu_{\amol} = \sum_{\amon} N_{\amol\amon}\delta\omega_{\amon}$ 
in the set of macroscopic chemical potential fields for molecules 
of different species, where $N_{\amol\amon}$ is the number of
$\amon$ monomers per molecule of type $\amol$. Such a homogeneous 
perturbation can have no effect upon monomer concentration fields 
in canonical ensemble, because the number of molecules of each 
type is constrained. It also can have no effects in either ensemble 
in an incompressible liquid with only one molecular species, such 
as a diblock copolymer melt, because the number of molecules per 
volume is then constrained by incompressibility.  

In either ensemble, we thus expect the matrix
$\srpa_{--}({\bf G},{\bf G}';\kv=0)$ in an incompressible diblock 
copolymer melt to have one vanishing eigenvalue, for which the only 
nonzero element of the corresponding eigenvector $f_{n}({\bf G};0)$ is 
the ${\bf G}=0$ element, corresponding to a homogeneous perturbation. 
This has long been known to be the case in the homogeneous phase of 
an incompressible diblock copolymer melt, for which the matrix
$\srpa_{--}({\bf G},{\bf G}';\kv=0)$ is diagonal, with a diagonal
element $S(\kv) = 0$ at $\kv={\bf G}={\bf G}'=0$.  We have 
confirmed that our numerical results for both $\sincmat_{--}$ and 
$\smat_{--}$ for ordered phases of diblock melts have this property. 
As in the disordered phase, we also find that the eigenvalue 
$\lambda_{n}(\kv)$ associated with one branch of the spectrum 
continuously approaches zero as $\kv \rightarrow 0$. To apply
Eq. (\ref{evalinv}) to a diblock copolymer melt at $\kv = 0$, 
one must thus identify and exclude this trivial zero mode. 

The spectrum of $\srpa_{\amon\bmon}({\bf G},{\bf G}';\kv=0)$ for 
a three dimensional structure generally also has three divergent 
eigenvalues, as a result of translational invariance. The corresponding 
eigenvectors are generators of rigid translations, which all have the 
form
\begin{equation}
  \delta \phi_{\amon}(\rv) = 
  \delta {\bf t} \cdot \nabla \phi_{\amon}(\rv)
  \label{dphi_translation}
\end{equation}
where $\delta {\bf t}$ is an infinitesimal rigid translation. Basis
vectors for the subspace of rigid translations may be obtained 
considering infinitesimal translations along three orthogonal 
directions. The inverse eigenvalue $\lambda_{n}^{-1}(\kv)$ associated
with these modes vanish because there is no free energy cost for rigid
translation of a crystal. In a band structure of $\lambda_{n}^{-1}(\kv)$ 
vs.  $\kv$ for a three dimensional structures, we thus find three 
``phonon-like" bands (one longitudinal and two transverse) in which 
the values of $\lambda_n^{-1}(\kv)$ approach zero as $k^{2}$ in 
the limit $\kv \rightarrow 0$. 

The fact that these ``phonon-like" eigenvalues of $\smat_{ij}$ diverge 
as $\kv \rightarrow 0$ does not cause any numerical problems
for diblock copolymer melts if we use Eq. (\ref{evalinv}) to calculate 
the eigenvalues of $\smat_{--}$. In this procedure, we numerically 
diagonalize the matrix $\sincmat_{--}$, for which the corresponding 
eigenvalues have finite values of $\tilde{\lambda}_n^{-1}(\kv=0) = 
\chi/2$. We find, however, that the our numerical results for 
$\lambda_{n}^{-1}(\kv=0)$ for these modes are very small only when 
the linear response is calculated for a well converged solution to the 
equilibrium SCFT, and only when the calculation is carried out with 
adequate spatial resolution. The behavior of these phonon-like modes 
thus provides a useful, and quite stringent, test of numerical 
accuracy.

\section{Space Group Symmetry}
\label{symmetry}
When calculating particular eigenvalues of the linear response matrix
$\srpa(\kv)$ at $\kv = 0$ or other special points in the Brillouin 
zone, it is sometimes possible to substantially reduce the cost of 
the eigenvalue calculation by making use of space group symmetry. 
As a simple example, if we know that the instability of a 
centrosymmetric structure is a result of an epitaxial instability 
towards another centrosymmetric structure, we expect the corresponding 
eigenvector of $\srpa(\kv)$ at $\kv=0$ to be even under inversion. To 
calculate the eigenvalue associated with such an instability, we may 
thus calculate a matrix representation of $\srpa(\kv)$ in the subspace 
of even functions by using a basis of cosine functions, rather than 
plane waves, and thereby reduce the number of basis functions by a 
factor of $2$ at a given spatial resolution. 

More generally, group theory can be used in the calculation of 
eigenvalues of $\srpa(\kv)$ at special points in the Brillouin zone 
in a manner very similar to what has long been used in the calculation
of eigenvalues of the Schroedinger equation in band structure 
calculations.\cite{bsw,tinkham_book} 
The starting point of the general analysis, in the present context, 
is the observation that the linear response operator $\smat(\kv)$ 
of a periodic structure is invariant under the symmetry elements 
of the so-called ``little group" $L(\kv)$ associated with crystal 
wavevector $\kv$.  In the case of eigenvectors at $\kv=0$, $L(\kv)$ 
is the same as the space group $G$ of the unperturbed crystal. More 
generally, $L(\kv)$ is the subgroup of the full space group $G$ 
containing symmetry elements that leave a plane wave 
$e^{i\kv\cdot\rv}$ of wavevector $\kv$ invariant. By arguments 
similar to those used to characterize the symmetry of eigenvectors 
of the Schroedinger equation at special $\kv$ points, it may 
be shown that (in the absence of accidental degeneracies) each 
eigenvalue of $\smat(\kv)$ may be associated with a specific 
irreducible representation of group $L(\kv)$.  Each such 
irreducible representation is associated with a subspace of 
functions that transform in a specified way under the action of 
the symmetry elements of $L(\kv)$. Eigenvectors with the symmetry
properties characteristic of a particular irreducible representation 
may be expanded using basis of functions that span the associated
subspace. 

As a trivial example, consider a one dimensional problem involving 
perturbations at $\kv = 0$ of a periodic structure that is 
symmetric under inversion (i.e., a centrosymmetric lamellar phase). 
The space group $G$ of the unperturbed crystal is the group $-1$, 
which contains only the identity element, $\rv \rightarrow \rv$, and 
the inversion element, $\rv \rightarrow -\rv$. At $\kv=0$, the 
relevant little group is the same as this full group. There are two 
possible irreducible representations of this group, for which the 
associated subspaces contain all functions $\psi(z)$ that are even 
under inversion, $\psi(z) = \psi(-z)$, or odd under inversion, 
$\psi(z) = -\psi(-z)$, respectively. Each eigenvector $\psi_{n}(z)$ 
must lie within one of these two subspaces, i.e., must be either 
even or odd. To calculate eigenvalues associated with the subset 
of eigenfunctions at $\kv=0$ that are even under inversion, we 
can use a cosine basis. To obtain eigenvalues associated with the 
remaining odd eigenfunctions, we can use a sine basis. Even if we 
require all of the eigenvalues, the size of the required secular 
matrices is reduced by considering even and odd subspaces separately, 
thereby block diagonalizing $\srpa$.

In general, to calculate an eigenvector or set of eigenvectors 
with a known symmetry, we may introduce a set of basis function with 
the desired symmetry. Let 
$f_{1}(\rv),f_{2}(\rv), \ldots , f_{M}(\rv)$ be a set of 
$M$ orthonormal basis functions that lie within the subspace of 
periodic functions associated with a given irreducible representation 
of the relevant little group $L(\kv)$.  Each of these basis functions 
is generally a superposition of plane waves with reciprocal lattice
wavevectors that are related to one another by the symmetry elements 
of group $L(\kv)$. The phase relationships among the coefficients of 
different plane waves within such a basis function are different in 
different irreducible representations. The symmetrized basis functions 
that we use to represent the solution of the unperturbed crystal, 
which are required to be invariant under all elements of the full 
space group $G$, are a special case of such basis functions, as are
cosine and sine functions.

To calculate the response within a subspace spanned by any such set 
of symmetry-adapted basis functions, we consider the response of an 
ideal gas to a perturbation of the form
\begin{equation}
    \delta\omega_{\bmon}(\rv) = e^{i\kv\cdot\rv}
    \sum_{\bbas}\delta\omega_{\bbas \bmon}f_{\bbas}(\rv)
    \quad.
\end{equation}
Such a perturbation is expected to yield a concentration perturbation 
$\delta\phi_{\abas}(\rv)$ with a periodic factor that can be expanded 
in terms of the same basis functions, with coefficients 
$\delta\phi_{\abas\amon}$. The linear response of the ideal gas in 
the subspace of interest may thus be characterized by a matrix
\begin{equation}
     \label{sid_basis}
     \delta \perden_{\abas\amon}
     = - \sid_{\abas\bbas,\amon\bmon}(\kv)
     \delta\perom_{\bbas\bmon}
\end{equation}
where $\sid_{\abas\bbas,\amon\bmon}(\kv)$ is a matrix representation 
of the ideal gas response $\sid(\kv)$ within the chosen subspace. 
The corresponding RPA response can be represented in this subspace 
by a matrix $\srpa_{\abas\bbas,\amon\bmon}(\kv)$ of the same form, using 
the same basis functions. The matrix equations that relate the ideal 
gas and RPA response matrices in this representation are identical 
to those obtained above for the special case of a plane wave basis, 
except for the replacement of summation over reciprocal vectors by 
summation over basis function indices $\abas$ and $\bbas$.

Our implementation of the linear response calculation allows for the
introduction of an arbitrary set of such basis functions. We have thus far
automated the generation of symmetry-adapted basis functions, however, only for
cases in which the required basis functions are invariant under all elements of
$L(\kv)$, or of a specified subgroup of $L(\kv)$. In these cases, the required
basis functions may be generated by the same algorithm as that used to generate
symmetry adapted basis functions for the solution of the unperturbed problem.
We have thus far actually used symmetry adapted basis functions, rather than
plane waves, only for the purpose of refining our results for the eigenvalues
of specific eigenvectors of $\srpa$ in the Gyroid phase, as discussed below.

When calculating eigenvalues for body-centered crystals using a simple cubic
cubic computational unit cell, we encountered a subtlety that is a result of
the translational symmetry that relates the two equivalent sublattices of 
the BCC structure. This is discussed in appendix \ref{app:Centering}.

\section{Algorithm and Efficiency}
\label{eff}
To calculate the eigenvalues of $\smat$ for an equilibrium 
structure of a diblock copolymer melt at a specific crystal 
wavevector $\kv$, we first calculate the equilibrium structure, 
and store the converged $\omega$ field. If the calculation will
use symmetry adapted basis functions, rather than a full plane 
wave basis, we next generate these functions. To calculate 
the spectrum of $\srpa$ in the invariant space spanned by the
chosen set of basis functions (which may be plane waves) we 
must then:
\begin{enumerate}
\item Use the perturbation theory of Sec. (\ref{sec:perturbation}) 
to calculate the ideal-gas response matrix
$\sid_{\amon\bmon}({\bf G},{\bf G}';\kv)$, in a plane 
wave basis, or
$\sid_{\abas\bbas,\amon\bmon}(\kv)$, in a basis of symmetry adapted 
functions.
\item Solve matrix Eq. (\ref{smm}) to obtain $\sincmat_{--}$. 
\item Diagonalize $\sincmat_{--}$.
\end{enumerate}
Once the eigenvalues of $\sincmat_{--}$ are known, Eq. (\ref{smminv}) 
may be used to obtain the corresponding eigenvalues of $\smat_{--}$. 

In step 1, we calculate $\sid_{\abas\bbas,\amon\bmon}$ using a truncated 
set of $M$ basis function, which may be either plane
waves or symmetry-adapted functions.  To do so, we calculate the 
perturbation $\delta \den_{\amon}$ in monomer concentration produced 
by perturbations of the form $\delta \omega_{\bbas\bmon}(\rv) \propto 
e^{i\kv\cdot\rv}f_{\bbas}(\rv)$, for every basis function $f_{\bbas}(\rv)$
in our basis set, for perturbations in both $\omega_{1}$ and 
$\omega_{2}$. This requires us to solve the perturbation theory 
for the ideal gas $2M$ times. Each such solution, which requires 
a calculation $\delta q$ and $\delta q^{\dagger}$, provides one 
row of the $2M \times 2M$ matrix $\sid_{\abas\bbas,\amon\bmon}$.

In the spectral method employed by Shi {\em et al.}\citep{shi_macro1},
${\cal O}(M^{3})$ floating point operations are required to calculate 
the response of an ideal gas to a single plane wave perturbation. 
The cost of a calculation of the entire matrix 
$\sid({\bf G},{\bf G}';\kv)$ for a single $\kv$ vector is thus 
${\cal O}(M^{4})$. The matrix operations required in steps (2) and (3) 
each require ${\cal O}(M^{3})$ operations. In the spectral method, 
the ${\cal O}(M^{4})$ cost of the calculation of the ideal gas 
susceptibility $\sid$ thus dominates the cost for large values 
of $M$. With this algorithm, Shi {\em et al.} \citep{shi_macro2} 
were limited to values of $M < 800$.

In our pseudo-spectral implementation, the calculation of the response 
to a single perturbation requires ${\cal O}(M_s M_g\ln{M_g})$ operations, 
where $M_g$ is the number of grid points, or plane waves, and $M_{s}$
is the number of discretized ``time-like'' steps along the chain. The 
cost of the calculation of the entire matrix 
$\sid({\bf G},{\bf G}';\kv)$ is thus ${\cal O}(M M_s M_g \ln{M_g})$. 
For plane wave calculations, $M = M_g$, and the cost of the calculation
is ${\cal O}(M_s M^2 \ln{M})$. We have used a time stepping algorithm, 
described in the appendix, that yields global errors of 
${\cal O}(\Delta s^{4})$, where $\Delta s$ is the contour length step 
size. With this algorithm, very high accuracy can be obtained with 
$M_{s} \simeq 10^{2}$. The pseudo-spectral algorithm for calculation 
of the ideal gas response function thus becomes much more efficient 
than the spectral method for large values of $M$. As a result, however, 
the ${\cal O}(M^{3})$ cost of the matrix inversion and diagonalization 
required in steps (2) and (3) will become the bottleneck for large $M$ 
in our algorithm.  A comparison of CPU times for the different parts 
of the calculation is given in Table \ref{Table:Cost}. The CPU time
required for steps 2 and 3 remains less than that of step 1 over the 
range grids reported here, but would begin to dominate for slightly 
larger values of $N$.

On a commodity personal computer, the calculation is also limited by 
the memory required in steps 2 and 3.  In our implementation, in which 
we have taken care to minimize memory usage, these matrix manipulations 
require storage of $2.5 M^{2}$ double precision real numbers for 
calculations involving centrosymmetric unperturbed crystals.

\begin{table}[h]
\center
\begin{tabular}{c|c|c|c|c|c}
\hline\hline
 $N$    &    $M$     &  time I.G.  & time L.A.   & memory I.G.  & memory L.A. \\ \hline
 8      &    512     &      2      &    0.2      &      19      &      4      \\ \hline
 12     &   1728     &     30      &    7        &      35      &     55      \\ \hline
 16     &   4096     &    219      &  106        &      61      &    320      \\ \hline
 20     &   8000     &   1053      &  771        &     107      &   1221      \\ 
\hline\hline
\end{tabular}
\caption{CPU time and memory costs for the calculation of $\srpa(\kv)$
and its eigenvalues for a Gyroid phase at a given crystal wavevector $\kv$, 
using a plane wave basis set on an $N \times N \times N$ grid, where $M = N^{3}$. 
Times are in minutes and memory in megabytes.  The time and memory required 
to calculate the ideal gas perturbation (step 1) are labeled time {I.G.} and 
memory I.G., respectively, while the costs of the linear algebra operations 
in steps 2 and 3 are labeled time {L.A.} and memory L.A. Each calculation was 
carried out on a single Athlon 2200 MP processor running at 1.7 GHz.}
\label{Table:Cost}
\end{table}

The efficiency of our calculation of limits of stability could be further 
improved in several ways that we have not yet explored:

A more efficient algorithm for calculating $\smat$ (step 2) for large 
values of $M$ might be obtained by replacing our direct matrix solution 
of the linearized SCF equation by an iterative calculation of the 
self-consistent field $\domvec$ produced by a given external field 
$\domvec^{\rm ext}$. The required iteration would be very similar to 
that which is normally used to solve the SCF equations for an equilibrium 
microstructure.  The cost of each iteration in such a method would be 
${\cal O}(M_{s}M\ln M)$. 

When only a limited number of low eigenvalues are of interest, as is
the case when determining limits of stability, efficient iterative 
methods could be used to solve the eigenvalue problem. Development 
of an efficient iterative solution of the linear SCF equations would 
provide a more efficient method of calculating the matrix-vector 
product $\drhovec = \smat\domvec^{\rm ext}$ for an arbitrary input 
vector $\domvec^{\rm ext}$. This could then be used as the inner 
operation of an iterative Krylov subspace method, such as the Lanczos 
method, to efficiently calculate the lowest eigenvalues of $\smat$. 

\section{Stability of Diblock Copolymer Phases}
\label{sec:diblock}

In this Section we present ``band diagrams'' for HEX, BCC and Gyroid phases 
in diblocks. The discussions for HEX and BCC phases are focused on the
interpretation of the degenerate unstable eigenmodes occurring at $\kv=0$.  
The discussion of Gyroid phase focuses on resolving the questions raised
by earlier studies of linear stability by Shi, Laradji and coworkers
\citep{shi_prl,shi_macro2} and by Matsen \cite{cg_matsen}.

\subsection{Hexagonally Ordered Cylinders}
First, we examine instabilities of the HEX phase toward BCC spheres, 
which occurs at $\kv \neq 0$, and towards a lamellar structure, which 
occurs at $\kv = 0$.  Experiments have shown that diblocks can undergo 
thermo-reversible transitions between cylinders to spheres
\citep{cst_fetters}.  

\begin{figure}
\begin{center}
\includegraphics[width=2.8in, height=!]{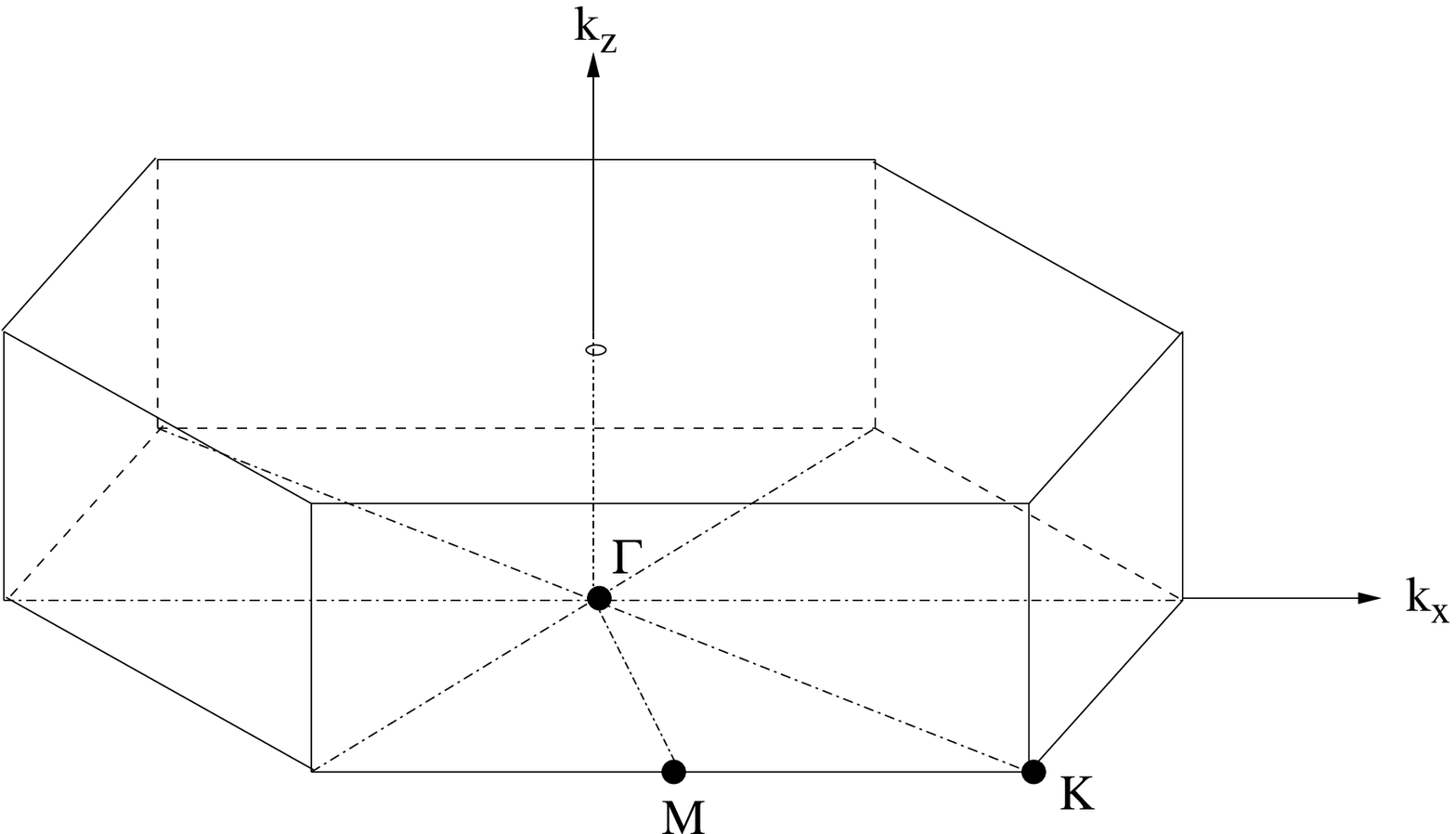}
\end{center}
\begin{center}
\includegraphics[width=3.0in, height=!]{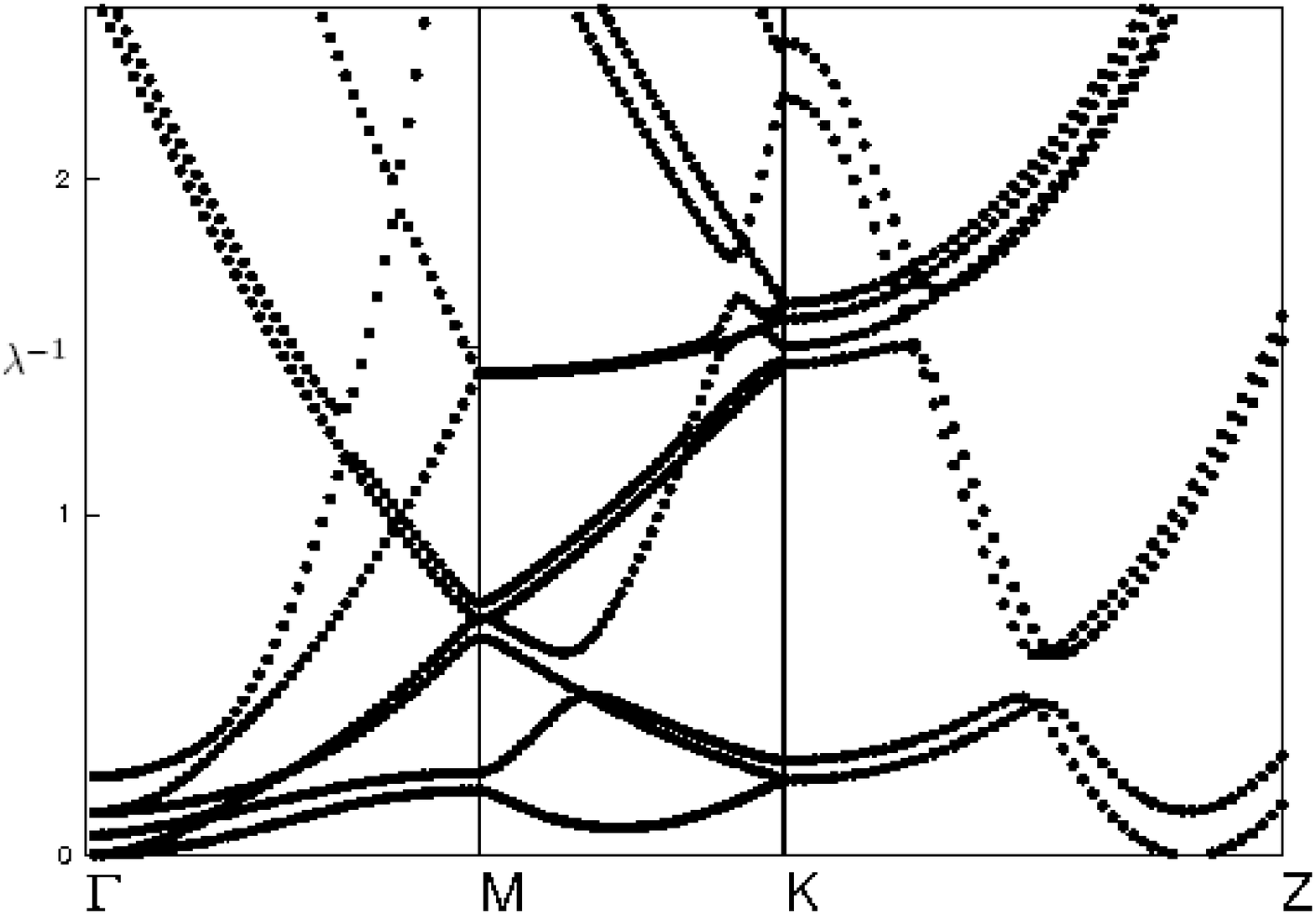}
\end{center}
\caption {The bands diagram of inverse eigenvalues of the
response matrix for the HEX phase.  The top subfigure shows 
the first Brillouin zone and the several points of high 
symmetry in the hexagonal structure \citep{jones_book}. 
The lower subfigure shows the band structure of the HEX 
phase at $f = 0.428$ and $\chi N = 10.9$, with the HEX-BCC 
instability at nonzero $\kv$. Subdiagrams labeled $\Gamma$-$M$ 
and $M$-$K$ in the band diagram show inverse eigenvalues 
calculated along corresponding lines in the plane $k_{z}=0$ 
in reciprocal space, while subdiagram $K$-$Z$ shows values
along a line constructed perpendicular to this plane through 
point $K$. 
\label{hexbcc} }
\end{figure}

\figref{hexbcc} shows the bands of inverse eigenvalues of the 
SCF response function for a HEX phase at $f = 0.428$ and 
$\chi N = 10.9$, which lies along the limit of stability of the 
HEX phase.  The band structure in this figure appears to agree with
that obtained by Shi {\em et al.} for the same choice of parameters. 
The instability at non-zero $\kv$ seen in this figure is an 
instability towards a BCC structure, which has been discussed by 
Shi {\em et al}.

\begin{figure}
\begin{center}
\includegraphics[width=3.0in,height=!]{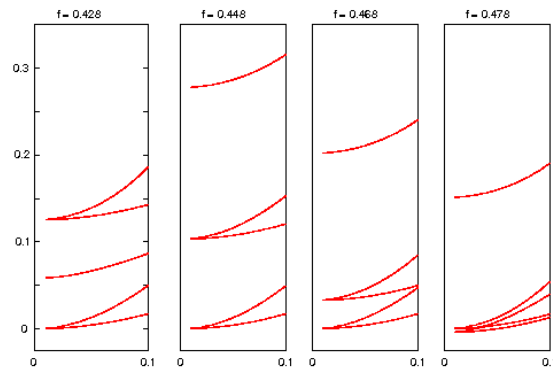}
\end{center}
\caption{The lowest few bands for the HEX phase at $\chi N = 10.9$ 
and at different compositions near the HEX to lamellar instability,
for small $\kv$ along the $\kv$-space line segment $\Gamma$-$M$. 
This instability occurs at $\kv=0$, at $f \simeq 0.478$.
\label{hexlam} }
\end{figure}

\figref{hexlam} shows the evolution of the first few bands in 
the HEX phase with changes in $f$ $\chi N = 10.9$ for a range of 
compositions near the limit of stability towards a lamellar phase. 
The structure becomes unstable at $f = 0.478$, when the inverse 
eigenvalues that are degenerate at $\kv=0$ simultaneously 
pass through zero at $\kv=0$. In addition to these unstable 
modes, this band diagram contains two phonon-like modes, for 
which $\lambda^{-1}_{n}(\kv=0) = 0$ for all choices of 
parameters. 

Examination of the two unstable eigenvectors at $\kv=0$ confirms 
that they have a structure consistent with an instability towards a 
lamellar phase.  Both of the degenerate unstable eigenmodes are found 
to be even under inversion.  We find that it is possible to construct 
three linear superpositions of these two eigenmodes such that each 
has a mirror symmetry through one of the three mirror planes of the 
hexagonal phase. 

A more detailed view may be obtained by considering the projections 
of these linear superpositions of the unstable modes onto the first 
``star" of reciprocal vectors ({\it i.e.}, the primary scattering 
peaks) for the HEX phase. Let these six primary reciprocal vectors 
be denoted 
${\bf G}_1, {\bf G}_2, {\bf G}_3, {\bf G}_4, {\bf G}_5, {\bf G}_6$,
with ${\bf G}_4 = -{\bf G}_1$, ${\bf G}_5 = -{\bf G}_2$, and 
${\bf G}_6 = -{\bf G}_3$. The projections of the unstable 
eigenmodes onto these vectors can be expressed in real space as 
a sum of three cosine functions with wave-vectors ${\bf G}_1$, 
${\bf G}_2$, and ${\bf G}_3$. The amplitudes of these cosine 
functions for the three superpositions discussed above, which
each exhibits a mirror plane, are ($1,-0.5,-0.5$), ($-0.5,1,-0.5$), 
and ($-0.5,-0.5,1$), respectively.  Each of these superpositions 
thus tends to increase the amplitude of one of the three cosine 
functions and decrease the amplitudes of the other two equally. 
This is what we expect for an epitaxial instability towards a 
lamellar phase that is aligned along any of three equivalent 
directions. 

\subsection{BCC Spheres}

Next, we examine the stability of the BCC spheres and the nature of 
the instability.  The instability has been considered previously by 
both Shi {\it et al.} \citep{shi_prl,shi_macro2} and Matsen 
\cite{cs_matsen}.

\begin{figure}
\begin{center}
\includegraphics[width=2.8in, height=!]{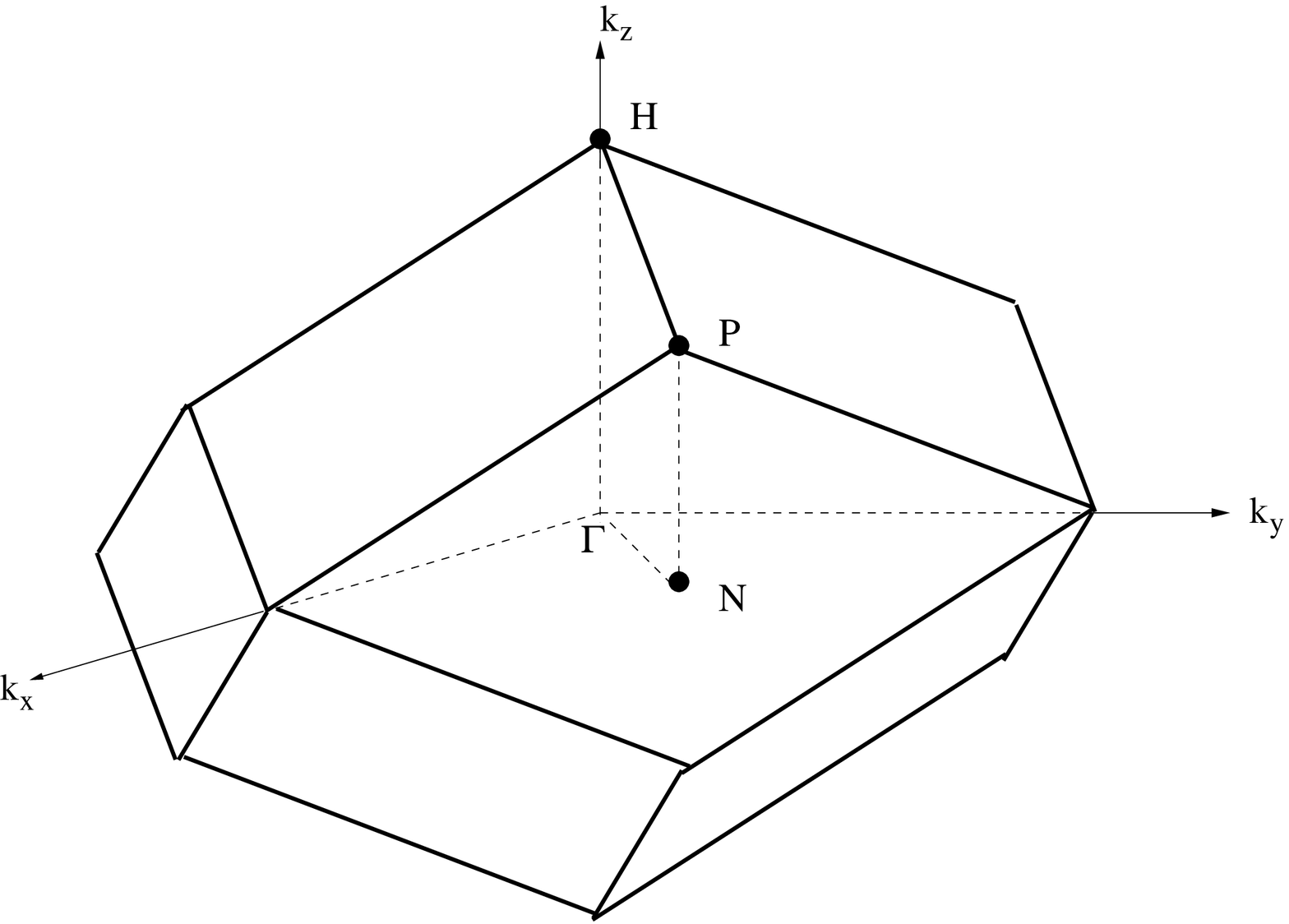}
\includegraphics[width=3.0in, height=!]{bcc_bands_f448_chi10.7.eps}
\end{center}
\caption{
Band diagrams of a stable BCC phase. The top subfigure shows 
the first Brillouin zone and labeled high symmetry points 
in $\kv$-space for a BCC crystal. \citep{jones_book} The lower 
subfigure shows the BCC bands at $\chi N=10.7$ and $f=0.448$, along 
several line segments in reciprocal space that connect the labeled 
pairs of points $\Gamma$-$H$, $H$-$P$, {\it etc.} 
The inverse eigenvalues smaller than 0.15 are plotted.
For this set of parameters the BCC structure is stable. At a slightly higher value 
of $\chi{N} = 10.8$ at the same composition the BCC phase becomes 
unstable, as shown in the \figref{bcc2}. \label{bcc1} }
\end{figure}

\begin{figure}
\begin{center}
\includegraphics[width=3.0in, height=!]{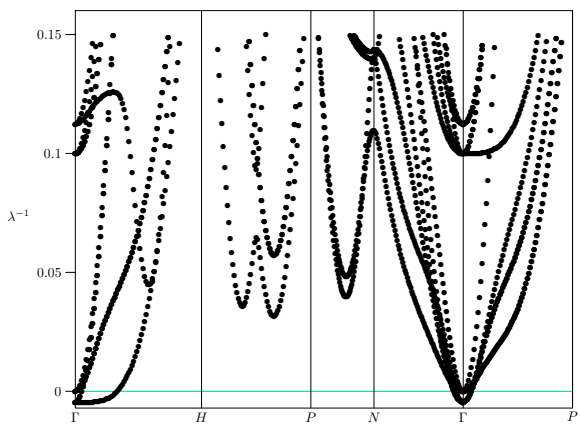}
\end{center}
\caption[Bands of an unstable BCC structure.]
{Band diagrams of a BCC phase at $\chi N=10.8$ and $f=0.448$. The inverse
eigenvalues smaller than 0.15 are plotted. The instability is seen to occur at
the $\Gamma$ point, {\it i.e.}, at $\kv=0$.\label{bcc2} }
\end{figure}

The BCC bands for $f = 0.448$ at two values of $\chi N$, 10.7 and
10.8, calculated using an $8\times8\times8$ grid, are shown
respectively in Figures~\ref{bcc1} and \ref{bcc2}.  
For $\chi{N} = 10.7$, the BCC structure is found stable for all 
$\kv$, whereas as $\chi{N} = 10.8$, the BCC equilibrium structure 
is unstable.  The instability in this structure sets in at 
$\kv = 0$ as seen in Figure \ref{bcc2}. The unstable eigenmode 
is triply degenerate at $\kv=0$. In addition to the unstable 
mode, the spectrum contains three phonon-like bands (one 
longitudinal and two transverse). The two transverse phonon bands 
are degenerate along the lines $\Gamma$-$P$ and $\Gamma$-$H$.

In this case, we find that it is possible to construct a linear 
superposition of the three degenerate unstable eigenmodes at 
$\kv = 0$ such that the resultant superposition has three 
fold symmetry about the $[111]$ axis. The resulting mode has 
positive and equal amplitudes for the 6 primary $\{011\}$ 
reciprocal lattice vectors that lie within a plane perpendicular 
to perpendicular $[111]$ (thus reinforcing these peaks), and 
negative amplitudes for the remaining 6 $\{011\}$ vectors (thus 
leading towards their extinction). This superposition corresponds 
to a modulation that leads towards the formation of hexagonal 
cylinders along the $[111]$ direction. Equivalent linear 
superpositions can be constructed for instabilities to cylindrical 
phases along the other $\langle 111\rangle$ directions.  We thus 
interpret the unstable mode as an epitaxial instability towards a 
hexagonal phase with cylinders along any of the $\langle 111 \rangle$ 
directions. 

This interpretation is consistent with the assumptions underlying 
Matsen's calculation, which could only describe instabilities of 
this type at $\kv=0$. Our interpretation is, however, different 
from that of Laradji {\em et al.} \citep{shi_macro2} who concluded 
that the instability of the BCC phase was an instability towards 
formation of a perforated lamellar structure with layers 
perpendicular to a $\{011\}$ direction. Laradji {\it et al.} did 
not report the fact that this unstable mode is degenerate. It 
appears likely to us that Laradji {\em et al.} overlooked the 
possibility of constructing an instability directly to the 
equilibrium HEX phase, rather than a metastable perforated 
lamellar phase, by a suitable linear superposition of the 3 
degenerate unstable eigenmodes.

\subsection{Gyroid Phase}
\label{gyrst}

Our results regarding the stability of the Gyroid phase 
require some discussion because earlier results by Laradji 
{\em et al.} have been controversial: 
Laradji, Shi {\it et al}. \citep{shi_macro1,shi_prl,shi_macro2} found 
that Gyroid phase was locally unstable at values of $\chi N \leq 12.0$,
within a range of the parameters $f$ and $\chi N$ in which the Gyroid 
phase was then believed to be the equilibrium phase. This result, if 
correct, would obviously be incompatible with the conclusion that the 
Gyroid phase was the global minimum in the free energy in this range 
of parameters. Our own interest in this question was initially raised 
by the discovery by our group \cite{tyler_prl,tyler_macro,ranjan} that 
an $Fddd$ orthorhombic network is actually the equilibrium structure 
in precisely the slice of parameter space (along the line of equal 
lamellar and HEX free energies) for which Laradji {\it et al.} 
reported the Gyroid to be unstable. This raised the question in 
our minds of whether Laradji {\it et al.}  might have identified an 
instability of the Gyroid phase towards an $Fddd$ phase of lower free 
energy. (This does not appear to be the case, as discussed below).

Matsen has argued instead that \citep{matsen_reply} 
Laradji {\it et al.}'s conclusion about the Gyroid phase was probably 
a result of numerical inaccuracy arising from the use of an 
insufficient number of plane waves.  Matsen has studied a pathway for 
epitaxial transformations between the Gyroid and a hexagonal cylinder 
phases in which the cylinders are aligned along the cubic [111] axis. 
\cite{cg_matsen} As part of this study, he examined the local 
stability of the Gyroid phase with respect to changes in the 
composition field that maintained the symmetries shared by the 
Gyroid and cylinder phase: He considered only perturbations of 
the Gyroid phase that retained the periodicity of the parent Gyroid 
phase (corresponding to instabilities at $\kv=0$), inversion 
symmetry, and three fold symmetry around the [111] axis.  Matsen 
found that the limit of stability of the Gyroid with respect to 
this type of instability lies near the line of equilibrium transitions 
between HEX and BCC phases, which is well beyond the calculated
line of equilibrium Gyroid-HEX transitions. He thus concluded 
that the Gyroid phase was locally stable with respect to this 
type of instability throughout, and well beyond, the region in 
which the Gyroid is known to have a lower SCF free energy than 
the HEX phase.

Matsen assumed throughout this controversy that the instability 
found by Laradji {\it et al.} must be an instability towards the 
HEX phase, of the type that he considered. This has remained less
clear to us, however, because Laradji {\it et al.} said nothing 
about the nature of the instability that they had identified, or 
even whether the instability occurred at zero or nonzero crystal 
wavevector. We have confirmed in private communications with Shi
that he and his coworkers did not ascertain the nature of the 
reported instability. One motivation for the work described here 
was thus to lay this question to rest, by repeating the calculation 
of the full response function without the restrictions on the 
symmetry or crystal wavevector of the perturbation, while using a 
significantly more efficient numerical method than that used by 
Laradji {\it et al.}.  We find, in agreement with Matsen, that 
the Gyroid phase is locally stable throughout region of parameter 
space in which it has a lower free energy than both the HEX and 
lamellar phases. 

\begin{figure}
\begin{center}
\includegraphics[width=3.0in,height=!]{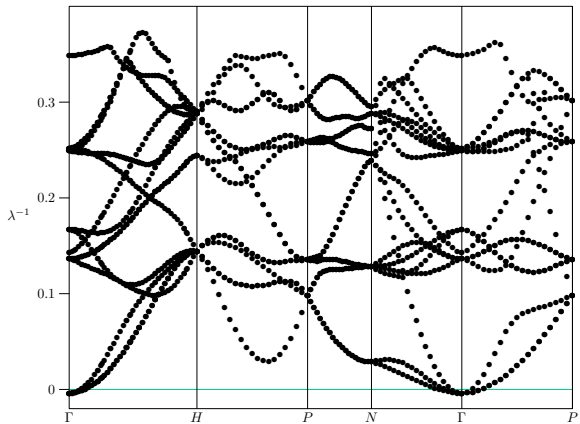}
\end{center}
\caption{
Band diagrams of a stable Gyroid phase at $f$=0.43, and $\chi{N}$=12, 
along the representative directions in $\kv$-space. The
first 16 inverse eigenvalues are plotted.
The labelling of special $\kv$-vectors is the same as those
for BCC structure in \figref{bcc1}.
\label{Gf43} }
\end{figure}

A band diagram for the Gyroid phase at $f_{A}=0.43$ and $\chi N=12$ 
is shown in \figref{Gf43}. These parameters correspond to those at 
which Laradji {\it et al.} concluded that the Gyroid was unstable, 
but that lie within the region in which Matsen and Schick found the 
Gyroid phase to be globally stable (i.e., to case 2 of Table 1 in 
Laradji {\it et al.} \cite{shi_macro2}). The eigenvalues in this
diagram were calculated using a $16 \times 16 \times 16$ grid, 
and a plane wave basis.

The lowest bands in this diagram are phonon-like modes. These have 
a small negative eigenvalue at the $\Gamma$ point ($\kv=0$) as a 
result of some remaining numerical error, which is not related to 
the physical instability of the structure.  We have confirmed that 
these three eigenmodes at $\kv=0$ correspond to rigid translations
by confirming that (to within small numerical errors) they span 
the same subspace as that spanned by Eq. (\ref{dphi_translation}) 
for the density modulation generated by arbitrary infinitesimal 
rigid translations. As a corollary of this, they are all found to
be odd under inversion. We show in Table \ref{Table:Convergence} 
that the associated value of $\lambda_n^{-1}(\kv=0)$ rapidly 
approaches zero as the grid is refined further.  Because all of
the other inverse eigenvalues in this diagram, are positive, we
conclude that the Gyroid is stable at this point in parameter
space. 

The physical instability of the Gyroid towards the HEX phase that
was considered by Matsen is associated with a three fold-degenerate 
eigenvalue at $\kv=0$ that is the next band up in this diagram. 
This set of three degenerate eigenmodes is found to have the same 
symmetry properties (i.e., the same irreducible representation) 
as those found for the corresponding instability of the BCC phase 
towards HEX: All three eigenvectors are even under inversion, and 
the space spanned by these eigenvectors contains an eigenvector 
with three-fold rotational symmetry about the $[111]$ axes, as 
well as equivalent eigenvectors with three-fold symmetry about 
each of the other other $\langle 111 \rangle$ axes. 

In light of the history of the problem, it is important for us 
to pay attention to questions of numerical convergence.
Table \ref{Table:Convergence} presents a study of the convergence 
with increasing spatial resolution of the eigenvalues associated
with both the rigid translation modes (labeled $T$) and these 
three dangerous modes for the $G \rightarrow H$ instability 
labeled ($H$) at $\kv=0$, for the same parameters of $f = 0.43$ 
and $\chi N = 12$.  Calculations with $N \times N \times N$ grids 
of $N$=8, 12, 16, and 20 have been carried out with a plane wave 
basis.  We can only use grids with $N$ being a multiple of 4 
because the grid must be invariant under the space group operations 
of group $Ia\bar{3}d$, which include diagonal (``$d$") glide 
operations that displace the structure by one-quarter of a unit 
cell. Calculations of the eigenvalue of the $H$ modes were also 
carried out on grids with $N=16$, $20$, $24$, and $28$ using 
basis functions with inversion symmetry and three fold rotational 
axis around the [111] axis, which require approximately 1/6 as 
many basis functions at each value of $N$.  Calculations of the 
eigenvalue associated with the rigid translation $T$ modes at 
$\kv=0$ were also carried out for $N$=16, 20, 24, and 28 using basis 
functions with three fold rotational symmetry about the [111] 
axis, with no imposed inversion symmetry. This leaves a single 
rigid translation mode that corresponds to a translation 
parallel to the [111] axis.  Calculations carried out with a 
plane wave basis and with symmetry adapted basis functions 
defined on the same grid, for $N=16$ and $N=20$, were found 
to yield corresponding eigenvalues that are identical to 
within the accuracy displayed in this table. Eigenvalues
calculated by different methods are thus not distinguished in 
the table.  The values given in the table, and those shown 
in Figs. 1-5, are all values of $\lambda_n^{-1}(\kv=0)$ for a 
diblock copolymer that has been non-dimensionalized by taking 
the reference volume equal to the chain volume, so that $N=1$.

For this set of parameters, we conclude that our results 
are adequately converged for $N \geq 16$, with errors of a 
few times $10^{-3}$ for $N=16$, but that qualitative errors
appear at $N=8$ and $N=12$.  We find that order of the 
eigenvalues associated with different eigenvectors at $\kv=0$
(which may be uniquely identified by their degeneracy and 
symmetry properties) is independent of $N$ for $N \geq 16$, 
but that the order is different for $N=8$ and $N=12$.  For 
$N=8$, at these set of parameters, we find a total of 12 
eigenvectors with negative values of $\lambda_n^{-1}(\kv=0)$, 
in several families of degenerate eigenvectors, among which 
are the 3 degenerate rigid translation/phonon modes. At $N=12$, 
only the 3 translation modes have negative eigenvalues. 
Accurate calculations for the Gyroid phase seem to require a 
substantially finer grid than required for the BCC phase, 
for which we obtain a comparable accuracy at similar values 
of $\chi N$ using $N = 8$. 

The number of plane waves used by Laradji {\it et al.} 
($M \leq 783$) corresponds most closely to that obtained 
with an $8 \times 8 \times 8$ grid ($M=512$), from which 
we obtain qualitatively incorrect results.  The conclusions 
of Laradji {\it et al.} regarding the stability of the 
Gyroid phase at this point in parameter space thus appear 
to be a result of inadequate spatial resolution.

\begin{table}[h]
\center
\begin{tabular}{c|c|c|c|c|c|c}
\hline\hline
    &    8      &     12      &    16       &     20      &      24     &    28       \\ \hline
 T  &  -0.2661  &  -3.605E-2  &  -4.422E-3  &  -3.791E-4  &  -3.828E-5  & -9.885E-7   \\ \hline
 H  &   0.1139  &   0.1140    &   0.1362    &   0.1372    &   0.1373    &  0.1373     \\ 
\hline\hline
\end{tabular}
\caption{$\lambda_n^{-1}$ at $\kv=0$ for the 3-fold degenerate rigid 
translation mode (labelled T) and the 3-fold degenerate dangerous mode for 
the $G\rightarrow\!H$ instability (labelled H) in the Gyroid phase at 
$f=0.43$ and $\chi N=12$, calculated using different $N \times N \times N$ 
grids, for $N=$8-28. }
\label{Table:Convergence}
\end{table}

\begin{figure}
  \begin{center}
  \includegraphics[width=3.00in, height=!]{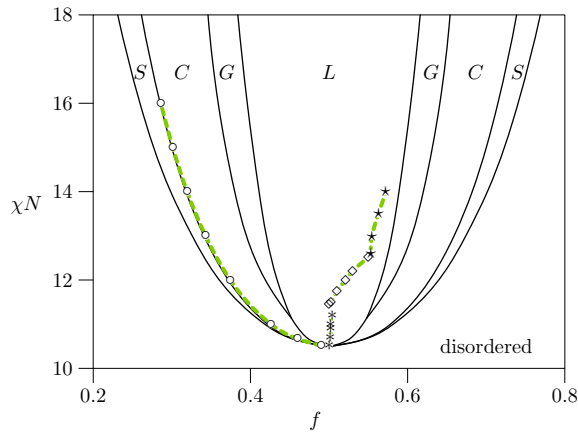}
  \end{center}
\caption{Stability limits of the Gyroid. The full lines are the 
equilibrium phase boundaries (excluding the $Fddd$ phase) and the 
open circles connected by dashed lines are various calculated limits 
of stability of the Gyroid structure. The position of the boundary 
at small values of $f$, which is an epitaxial ($\kv=0$) instability 
towards a HEX phase, is found to agree well with Matsen's results 
\citep{cg_matsen}.  Another eigenmode at $\kv \neq 0$ is found
to become unstable along a line that would be indistinguishable 
from this one at the scale of this figure.  The other symbols 
$\star$, $\diamond$, $\ast$ and the connecting dashed lines indicate 
the other types of unstable modes, as discussed in the text.
\label{stgyr} }
\end{figure}

We next considered the limits of stability of the Gyroid phase with 
respect to the epitaxial instability that was considered previously 
by Matsen. To do so, we calculated eigenvalues at $\kv=0$ over a 
range of values of $f_{A}$ for several values of $\chi N$.  The 
three-fold degenerate inverse eigenvalue $\lambda_n^{-1}(\kv=0)$ 
associated with the epitaxial $G \rightarrow H$ instability was 
found to decrease and pass through zero with decreasing $f_{A}$,
at each value of $\chi N$, and to be the first inverse eigenvalue 
at $\kv=0$ to become unstable.
The resulting limit of stability with respect to this eigenmode is 
shown in \figref{stgyr} by the dashed line with open circles. Our
results for this line of instabilities, which lies very close to 
the equilibrium order-order transition between BCC spheres and 
hexagonal cylinders, agrees very well with Matsen's result for the 
same eigenmode \citep{cg_matsen}.

\begin{figure}
\begin{center}
\includegraphics[width=3.0in,height=!]{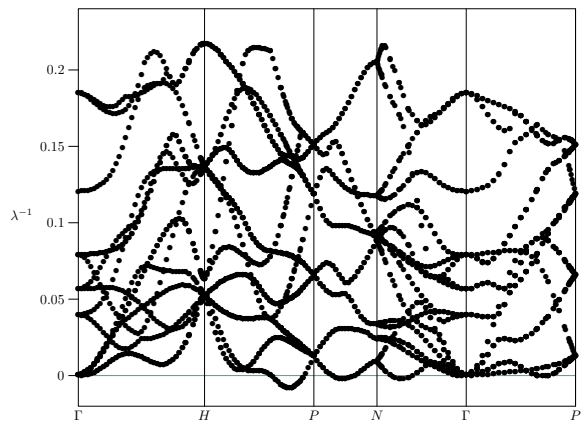}
\end{center}
\caption{
Band diagrams for the Gyroid phase at $f=0.3745$, and 
$\chi{N}=12$, along the representative directions in $\kv$-space. 
The first 16 inverse eigenvalues are plotted. The labeling of 
special $\kv$-vectors is the same as in \figref{Gf43}.
\label{Gf3745} }
\end{figure}

To check whether this instability at $\kv = 0$ is preempted by another
instability at some $\kv \neq 0$, we then calculated another band diagram at a
point $f=0.3745$, $\chi N=12$ along the proposed limit of stability line, at
which the value of $\lambda_{n}^{-1}(\kv=0)$ corresponding to the epitaxial $G
\rightarrow H$ instability exactly vanishes. The results are shown in
\figref{Gf3745}. At these conditions, 6 eigenvectors have nearly vanishing
eigenvalues at $\kv=0$ (the $\Gamma$ point), corresponding to the 3 rigid
translation modes and the three $G \rightarrow H$ modes. For this pair of
parameters, however, we also found very small negative values of
$\lambda^{-1}_{n}(\kv)$ at several points near the $P$ and $N$ points: Unstable
modes were found along the $HP$(1), $PN$(2), $N\Gamma$(1) and $\Gamma P$ (2)
directions. (The numbers in parentheses indicate the degree of degeneracy of
the relevant bands along these high-symmetry lines.) The most negative
eigenmode in this diagram, which we assume to correspond to the true limit of
stability, lies along the $HP$ direction.  By calculating eigenvalues at
several nearby points along two lines constructed through this point along
directions perpendicular to the $HP$ line, we have confirmed that this is a
local minimum of $\lambda^{-1}_{n}(\kv)$ for the lowest band in the
three-dimensional band diagram.

To investigate further, we thus ran a series of calculations of the band
diagram along the $HP$ line for several of values of $f$ near the previously
calculated limit of stability of the epitaxial mode, for integer values of
$\chi N$ = 11-16.  At each value of $\chi N$, we found a minimum value of
$\lambda^{-1}(\kv)$ vs $\kv$ at approximately the same wavevector $\kv$, which
is displaced from the $P$ point by a fraction $0.25$ of the distance between
$P$ and $H$ for $\chi N=12$ and which becomes closer to the $H$ point as the
value of $\chi N$ increases. At each value of $\chi N$, we found that this
minimum value of $\lambda^{-1}(\kv)$ at $\kv \neq 0$ passed through zero at a
value of $f$ that is within approximately $10^{-3}$ of that at which the
epitaxial $\kv=0$ modes become unstable.  For example, at $\chi N=12$, we find
an instability at $\kv \neq 0$ at $f=0.3757$, vs. $f=0.3745$ for the
instability at $\kv = 0$. Differences in the two critical values of $f$ are
equally small for the other values of $\chi N$. We are not confident of our
ability to accurately resolve such small differences in the critical values of
$f$ with the $16 \times 16 \times 16$ grid used in these calculations.  We thus
conclude only that this unexpected incommensurate instability competes
extremely closely with the epitaxial instability considered by Matsen, and may
well preempt it. 

We do not have a physical interpretation of the nature 
of this incommensurate instability, or toward what type 
of structure it might lead. We note, however, that the
identification of such an instability at $\kv \neq 0$ 
could not have been accomplished by Matsen's approach, 
which allowed him to consider only stability with respect 
to $\kv=0$ eigenmodes with a specified residual symmetry: 
It required the development of an efficient method of 
calculating the full linear response at arbitrary $\kv$. 

In \figref{stgyr} we have also shown three other limits of stability of 
this Gyroid structure at larger values of $f$. The branch labeled with 
$\diamond$ is obtained by tracing the equilibrium unit cell size $a$ of 
the Gyroid structure, and is characterized by a rapid contraction of 
the unit cell with increasing $f$: As this line is approached from the 
left, the derivative $\partial a/\partial f$ appears to diverge to 
$-\infty$. The branch labeled with $\star$ is obtained 
from perturbation calculation at $\kv=0$ and is doubly degenerate, with 
even eigenfunctions. It is often found to be accompanied by one of the 
other two unstable modes not shown in the figure, also occurring at 
$\kv=0$. One of these two modes is non-degenerate, for which the 
eigenfunction is odd under inversion, and has three fold rotational 
symmetry along $\langle111\rangle$ directions. Another one is triply 
degenerate, with eigenfunctions that are even. The branch labeled with 
$\ast$ is also obtained from perturbation calculation at $\kv=0$ and 
is the result of a triply degenerate instability that appears to 
cross the $G\rightarrow H$ instability in the vicinity of the critical 
point. Along this high-$f$ boundary of the Gyroid phase, where one
might expect a transition to a lamellar phase, we made no attempt to 
look for yet more instabilities at $\kv \neq 0$. 

\section{Conclusions}
\label{conclusions}
We have developed a pseudo-spectral algorithm for calculating the 
linear susceptibility of the ordered phases in block copolymeric 
melts that is significantly more efficient than that employed in 
earlier work by Shi and coworkers. We have used the new algorithm 
to re-examine the limits of stability of several ordered phases 
in diblock copolymers, and have resolved a controversy regarding 
the local stability of the Gyroid phase. We have also identified
an unexpected instability of the Gyroid phase at a nonzero $\kv$
vector along the zone edge, which competes very closely with the
epitaxial $G \rightarrow H$ transition considered previously by
Matsen. 

{\it Acknowledgments}: This work was supported primarily by the 
MRSEC Program of the National Science Foundation under Award Number
DMR-0212302, using computer resources provided by MRSEC and by the
University of Minnesota Supercomputer Institute. We are also grateful 
for helpful conversations with both An-Chang Shi and Mark Matsen. 

\appendix
\section{Integration Algorithm}
\label{idgas}
To calculate response of an ideal gas to a specified perturbation, 
we must numerically solve Equation (\ref{per_de}) for $\delta q({\bf
r},s)$.  We do so by discretizing the ``time-like" variable $s$ into
steps of equal size $\Delta s$ and numerically integrating the partial
differential equation. To carry out the integration, we use a
pseudo-spectral algorithm closely analogous to one that was introduced
by Rasmussen and Kalosakas \citep{kalosakas} as an algorithm for 
solving the unperturbed MDE that is solved to describe unperturbed
equilibriums state. We have combined a pseudo-spectral algorithm 
with Richardson extrapolation to obtain solutions with errors of 
${\cal O}(\Delta s^{4})$.

\subsection{Unperturbed MDE}
In this subsection, we review the Rasmussen-Kalosakas (RK) algorithm 
for the solution of the unperturbed for $q(\rv,s)$, and present 
an extrapolation method that we use to improve the accuracy of this
algorithm. The RK algorithm is based upon a representation of 
$q(\rv,s)$ at each value of $s$ on a regular spatial grid, and 
the use of a Fast Fourier Transform (FFT) to transform between 
real- and Fourier-space representations. Given a solution 
$q(\rv,s_{n})$ at $s=s_{n}$, the value at $s_{n+1} = 
s_{n}+\Delta s$ may be expressed formally as a product
\begin{equation}
   q(s_{n+1}) = e^{-\Delta s \; H}q(s_{n})
\end{equation}
where $q(s_{n})$ represents the function $q(\rv,s_{n})$, and
where $e^{-\Delta s \; H}$ is an exponential operator. In the RK 
algorithm, the propagator is approximated by
\begin{equation}
    \exp(-H \Delta s) \simeq
    e^{ -\frac{1}{2}\Delta s \; \omega }
    e^{ \Delta s \; \frac{b^2}{6}\nabla^{2} }
    e^{ -\frac{1}{2}\Delta s \; \omega }
    \quad.
\end{equation}
The product $e^{-H\Delta s}q(s_{n})$ is then evaluated by:
\begin{enumerate}
\item Evaluating a product 
\begin{equation}
  q_{n}^{(+)}(\rv)
  = e^{ -\frac{\Delta s}{2} \omega(\rv) }
    q(\rv,s_{n})
\end{equation}
at regularly spaced grid points,
\item Applying a FFT to obtain $q_n^{+}(\kv)$ and using the 
Fourier representation to evaluate 
\begin{equation}
  q^{(-)}_{n+1}(\kv)
  = e^{ -\Delta s\frac{b^{2}}{6}|\kv|^{2} }
  q_{n}^{(+)}(\kv)
\end{equation}
\item Applying an inverse FFT to obtain 
$q_{n+1}^{(-)}(\rv)$ and again evaluating a product
\begin{equation}
  q(\rv,s_{n+1})
  = e^{ -\frac{\Delta s}{2} \omega(\rv) }
  q^{(-)}_{n+1}(\rv)  \quad,
\end{equation}
on the real-space grid. 
\end{enumerate}
This algorithm yields a solution with local errors of ${\cal O}
(\Delta s^{3})$, or global errors of ${\cal O}(\Delta s^{2})$

To improve the accuracy of the solution we have used an extrapolation
scheme in which we calculate each time step using two different values
of $\Delta s$, which differ by a factor of 2, and then extrapolating 
to $\Delta s = 0$ to obtain the next value. Given $q(s_{n})$, we first
calculate a function $q(s_{n+1}; \Delta s)$ by applying the RK 
algorithm once with a time step $\Delta s$. We then 
calculate a function $q(s_{n+1}; \Delta s/2)$ by applying the above 
algorithm twice, using a step size $\Delta s/2$.  The final value of 
$q(s_{n+1})$, which is used as the starting point for the next step, 
is obtained from the extrapolation
\begin{equation}
    \label{RK-extrap}
    q(s_{n+1}) = [ \; 4 q(s_{n+1};\Delta s/2) - q(s_{n+1};\Delta s)\; ]/3
   \quad,
\end{equation}
which is designed to cancel the accumulating errors of order $(\Delta s)^{2}$. 

This extrapolation scheme for the unperturbed MDE was originally implemented in
the pseudo-spectral version of our SCFT code. When designed, it was
expected to remove global errors of order ${\cal O}(\Delta s^{2})$, and to
leave errors of ${\cal O}(\Delta s^{3})$. When the algorithm was tested,
however, it was found to yield a global error that decreased with $\Delta s$ as
$(\Delta s)^{4}$. We now believe that this behavior is a result of a special
property of reversible integration algorithms \cite{odebook}.  An algorithm is
said to be reversible if one forward propagation of the solution $q(s_{n})$,
followed by a backward propagation with the same step size, can exactly recover
the starting point $q(s_n)$.  The RK algorithm is reversible in this sense. It
is known \cite{odebook} that the Taylor series expansion of the global error
produced by a reversible discrete integrator for any system of first order
differential equations contains only even powers of $\Delta s$.  As a result,
reversible integration algorithms always exhibit the behavior we observed for
the extrapolated RK algorithm: An extrapolation that is designed to decrease
the global errors from ${\cal O}(\Delta s)^{2n}$ to order $(\Delta s)^{2n+1}$
generally yields a solution with global errors of order $(\Delta s)^{2n+2}$.

\subsection{Perturbed MDE}
We now describe the algorithm used to solve Eq. (\ref{per_de}), and
the conjugate equation for $\delta q^{\dagger}$.  
A formal solution of the Equation~\ref{per_de} over a single step of
length $\Delta s$ can be written as an integral:
\begin{equation}
  \label{eqpert}
  \delta\perq(s_{n+1}) = G_{\kv}(\Delta s)\delta\perq(s_{n}) +
  \int_{0}^{\Delta s} ds' G_{\kv}(s')f(s_{n+1}-s')
\end{equation}
where 
\begin{equation}
f(\rv,s) = -\delta\perom(\rv;\kv)q_{0}(\rv,s)
\end{equation} 
is the inhomogeneous term in this linear PDE, and where
$G_{\kv}(s) \equiv e^{-s H_{\kv}}$ is the propagator for
perturbations of crystal wavevector $\kv$.

\begin{figure}
  \begin{center}
  \includegraphics[scale=0.25,angle=-90]{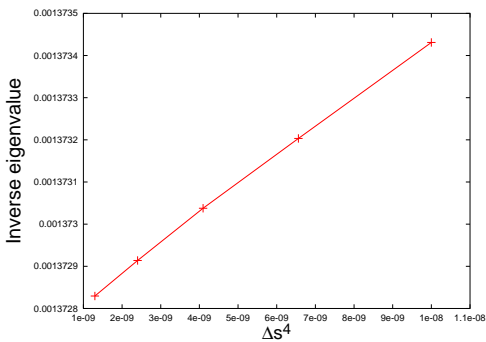}
  \end{center}
  \caption[First eigenvalue of the inverse RPA response function for
  a lamellar phase vs. $\Delta{s}^4$, when first order extrapolation 
  is used]{First inverse eigenvalue of the response function $\smat$
  a lamellar phase vs. $\Delta{s}^4$ when the perturbation theory is 
  evaluated using the extrapolation scheme of Eq. \ref{RK-extrap}. 
  \label{ext1} }
\end{figure}

Our integration scheme (prior to Richardson extrapolation) is to 
take
\begin{equation}
   \label{pert_algorithm}
   \delta\perq(s_{n+1}) = G(\Delta s) \delta\perq(s_{n}) +
                          G(\Delta s/2) \bar{f}(s_{n}) \Delta s
\end{equation}
where
\begin{equation}
   \bar{f}(s_{n}) \equiv [ f(s_{n+1}) + f(s_{n}) ] / 2
\end{equation}
while using the RK approximation for propagation by both 
$G(\Delta s)$ and $G(\Delta s/2)$. This algorithm, like the RK
algorithm, yields global errors of ${\cal O}(\Delta s^{2})$. 
Also, like the RK algorithm, it is reversible.  We thus use the 
same extrapolation scheme for this algorithm as that given in 
Eq. (\ref{RK-extrap}) for the RK algorithm. After $\delta q$ and 
$\delta q^{\dagger}$ are obtained by this method, the integral 
with respect to $s$ in Eq. (\ref{delrho}) is evaluated using 
Simpson's method.  The extrapolated solution for $\delta\den$, 
and thus for the matrix $\smat$, have errors of 
${\cal O}(\Delta s^{4})$, as demonstrated in Fig. \ref{ext1}.

\section{Body- and Face-Centered Lattices}
\label{app:Centering}
Here, we explain an issue that we encountered when calculating band
diagrams for BCC and Gyroid phases using a non-primitive cubic unit 
cell. Analogous issues can arise whenever a body- or face-centered
crystal is treated with a non-primitive computational unit cell. 

Band diagrams for the BCC and Gyroid phases, can be calculated using 
either a cubic unit cell with orthogonal axes, or a primitive unit 
cell with non-orthogonal axes, which has half the volume of the cubic
unit cell. (The Gyroid structure is based on a BCC lattice). We have 
used a simple cubic computational unit cell with cell size $a$, and 
discretized this with a simple cubic FFT grid.  The reciprocal lattice 
associated with this computational unit cell thus includes wavevectors 
that are not part of the reciprocal lattice of the BCC Bravais lattice, 
which is an FCC lattice in $k$-space. The first Brillouin zone (FBZ) 
associated with the simple cubic unit cell 
($|k_x|, |k_y|, |k_z| < \pi/2$) has half the volume in $k$-space as 
the FBZ for the BCC unit cell. 

If the algorithm described in this paper is applied to a BCC structure
using a simple cubic cell, without explicitly accounting for the centering 
symmetry of the unperturbed structure, the list of eigenvalues obtained 
at a specified $\kv$ in the FBZ of the simple cubic cell include both 
those associated with $\kv$ and those associated with another wavevector 
$\kv'=\kv+\Gv$ that differs from $\kv$ by a lattice vector 
$\Gv=(\pm 1,0,0)2\pi/a$, $(0,\pm 1,0)2\pi/a$, or $(0,0,\pm 1)2\pi/a$ 
that is part of the simple cubic reciprocal lattice, but not part of 
the reciprocal lattice of BCC. The vectors $\kv$ and $\kv'$ are thus 
equivalent from the point of view of the simple cubic lattice but 
inequivalent from the point of view of the BCC lattice.  Correspondingly, 
if $\kv$ lies in the FBZ of the simple cubic cell, then $\kv'$ generally 
lies within the FBZ of the BCC unit cell but the outside the smaller FBZ 
of the simple cubic unit cell. The result is a ``folding'' of the FBZ of 
the BCC unit cell into the smaller FBZ of the simple cubic computational
unit cell.

The problem could be avoided either by using
a primitive BCC unit cell from the outset, or by using only the 
reciprocal vectors of the BCC unit cell in the calculation of 
$S_{ij}(\Gv,\Gv';\kv)$. What we have actually done is to use our 
machinery for generating symmetrized basis functions to generate 
basis functions that are invariant under a subgroup of $L(\kv)$ 
group that, at a minimum, includes the identity 
$\rv \rightarrow \rv$ and the body-centering translational 
symmetry $\rv \rightarrow \rv + (1,1,1)a/2$. This guarantees 
that we will only obtain eigenfunctions of the Bloch form 
$e^{i\kv\cdot\rv}\psi_{n}(\rv)$ in which $\psi_{n}(\rv)$ has 
the periodicity of the BCC lattice, and not just the periodicity
of the larger cubic cell. If only these two symmetry elements 
are included, the algorithm automatically generates basis 
functions that are simply plane waves with wavevectors that 
belong to the reciprocal lattice for BCC, while discarding
the remaining ``extinct'' reciprocal lattice vectors of the 
simple cubic lattice.  The advantage of this approach is its 
generality: It allows the use of either primitive or non-primitive 
unit cells, generalizes immediately to the treatment of other 
kinds of body- and face-centered crystals, and does not require 
the explicit addition of special extinction conditions to our
algorithm.


\begin{thebibliography}{27}
\expandafter\ifx\csname natexlab\endcsname\relax\def\natexlab#1{#1}\fi
\expandafter\ifx\csname bibnamefont\endcsname\relax
  \def\bibnamefont#1{#1}\fi
\expandafter\ifx\csname bibfnamefont\endcsname\relax
  \def\bibfnamefont#1{#1}\fi
\expandafter\ifx\csname citenamefont\endcsname\relax
  \def\citenamefont#1{#1}\fi
\expandafter\ifx\csname url\endcsname\relax
  \def\url#1{\texttt{#1}}\fi
\expandafter\ifx\csname urlprefix\endcsname\relax\def\urlprefix{URL }\fi
\providecommand{\bibinfo}[2]{#2}
\providecommand{\eprint}[2][]{\url{#2}}

\bibitem[{\citenamefont{de~Gennes}(1979)}]{dgbook}
\bibinfo{author}{\bibfnamefont{P.-G.} \bibnamefont{de~Gennes}},
  \emph{\bibinfo{title}{Scaling Concepts in Polymer Physics}}
  (\bibinfo{publisher}{Cornell University Press}, \bibinfo{year}{1979}).

\bibitem[{\citenamefont{Nozieres and Pines}(1958{\natexlab{a}})}]{pines1}
\bibinfo{author}{\bibfnamefont{P.}~\bibnamefont{Nozieres}} \bibnamefont{and}
  \bibinfo{author}{\bibfnamefont{D.}~\bibnamefont{Pines}},
  \bibinfo{journal}{Physical Review} \textbf{\bibinfo{volume}{109}},
  \bibinfo{pages}{741} (\bibinfo{year}{1958}{\natexlab{a}}).

\bibitem[{\citenamefont{Nozieres and Pines}(1958{\natexlab{b}})}]{pines2}
\bibinfo{author}{\bibfnamefont{P.}~\bibnamefont{Nozieres}} \bibnamefont{and}
  \bibinfo{author}{\bibfnamefont{D.}~\bibnamefont{Pines}},
  \bibinfo{journal}{Physical Review} \textbf{\bibinfo{volume}{109}},
  \bibinfo{pages}{762} (\bibinfo{year}{1958}{\natexlab{b}}).

\bibitem[{\citenamefont{Pines and Nozieres}(1989)}]{pinesnozieres}
\bibinfo{author}{\bibfnamefont{D.}~\bibnamefont{Pines}} \bibnamefont{and}
  \bibinfo{author}{\bibfnamefont{P.}~\bibnamefont{Nozieres}},
  \emph{\bibinfo{title}{Theory of Quantum Liquids}}
  (\bibinfo{publisher}{Addison Wesley Publishing Company},
  \bibinfo{year}{1989}).

\bibitem[{\citenamefont{Ashcroft and Mermin}(1976)}]{ashmer}
\bibinfo{author}{\bibfnamefont{N.~W.} \bibnamefont{Ashcroft}} \bibnamefont{and}
  \bibinfo{author}{\bibfnamefont{N.~D.} \bibnamefont{Mermin}},
  \emph{\bibinfo{title}{Solid State Physics}} (\bibinfo{publisher}{Saunders
  College Publishers}, \bibinfo{year}{1976}).

\bibitem[{\citenamefont{Ziman}(1972)}]{ziman_book}
\bibinfo{author}{\bibfnamefont{J.~M.} \bibnamefont{Ziman}},
  \emph{\bibinfo{title}{Principles of the Theory of Solids}}
  (\bibinfo{publisher}{Cambridge University Press}, \bibinfo{year}{1972}).

\bibitem[{\citenamefont{Leibler}(1980)}]{leibler}
\bibinfo{author}{\bibfnamefont{L.}~\bibnamefont{Leibler}},
  \bibinfo{journal}{Macromolecules} \textbf{\bibinfo{volume}{13}},
  \bibinfo{pages}{1602} (\bibinfo{year}{1980}).

\bibitem[{\citenamefont{Shi et~al.}(1996)\citenamefont{Shi, Noolandi, , and
  Desai}}]{shi_macro1}
\bibinfo{author}{\bibfnamefont{A.-C.} \bibnamefont{Shi}},
  \bibinfo{author}{\bibfnamefont{J.}~\bibnamefont{Noolandi}}, ,
  \bibnamefont{and} \bibinfo{author}{\bibfnamefont{R.~C.} \bibnamefont{Desai}},
  \bibinfo{journal}{Macromolecules} \textbf{\bibinfo{volume}{29}},
  \bibinfo{pages}{6487} (\bibinfo{year}{1996}).

\bibitem[{\citenamefont{Laradji
  et~al.}(1997{\natexlab{a}})\citenamefont{Laradji, Shi, Desai, and
  Noolandi}}]{shi_prl}
\bibinfo{author}{\bibfnamefont{M.}~\bibnamefont{Laradji}},
  \bibinfo{author}{\bibfnamefont{A.-C.} \bibnamefont{Shi}},
  \bibinfo{author}{\bibfnamefont{R.~C.} \bibnamefont{Desai}}, \bibnamefont{and}
  \bibinfo{author}{\bibfnamefont{J.}~\bibnamefont{Noolandi}},
  \bibinfo{journal}{Physical Review Letters} \textbf{\bibinfo{volume}{78}},
  \bibinfo{pages}{2577} (\bibinfo{year}{1997}{\natexlab{a}}).

\bibitem[{\citenamefont{Laradji
  et~al.}(1997{\natexlab{b}})\citenamefont{Laradji, Shi, Desai, and
  Noolandi}}]{shi_macro2}
\bibinfo{author}{\bibfnamefont{M.}~\bibnamefont{Laradji}},
  \bibinfo{author}{\bibfnamefont{A.-C.} \bibnamefont{Shi}},
  \bibinfo{author}{\bibfnamefont{R.~C.} \bibnamefont{Desai}}, \bibnamefont{and}
  \bibinfo{author}{\bibfnamefont{J.}~\bibnamefont{Noolandi}},
  \bibinfo{journal}{Macromolecules} \textbf{\bibinfo{volume}{30}},
  \bibinfo{pages}{3242} (\bibinfo{year}{1997}{\natexlab{b}}).

\bibitem[{\citenamefont{Matsen and Schick}(1994)}]{matsen_schick}
\bibinfo{author}{\bibfnamefont{M.~W.} \bibnamefont{Matsen}} \bibnamefont{and}
  \bibinfo{author}{\bibfnamefont{M.}~\bibnamefont{Schick}},
  \bibinfo{journal}{Physical Review Letters} \textbf{\bibinfo{volume}{72}},
  \bibinfo{pages}{2660} (\bibinfo{year}{1994}).

\bibitem[{\citenamefont{Matsen}(1998{\natexlab{a}})}]{cg_matsen}
\bibinfo{author}{\bibfnamefont{M.~W.} \bibnamefont{Matsen}},
  \bibinfo{journal}{Physical Review Letters} \textbf{\bibinfo{volume}{80}},
  \bibinfo{pages}{440} (\bibinfo{year}{1998}{\natexlab{a}}).

\bibitem[{\citenamefont{Matsen}(2001)}]{cs_matsen}
\bibinfo{author}{\bibfnamefont{M.~W.} \bibnamefont{Matsen}},
  \bibinfo{journal}{Journal of Chemical Physics}
  \textbf{\bibinfo{volume}{114}}, \bibinfo{pages}{8165} (\bibinfo{year}{2001}).

\bibitem[{\citenamefont{Rasmussen and Kalosakas}(2002)}]{kalosakas}
\bibinfo{author}{\bibfnamefont{K.~O.} \bibnamefont{Rasmussen}}
  \bibnamefont{and}
  \bibinfo{author}{\bibfnamefont{G.}~\bibnamefont{Kalosakas}},
  \bibinfo{journal}{Journal of Polymer Sci. B} \textbf{\bibinfo{volume}{40}},
  \bibinfo{pages}{1777} (\bibinfo{year}{2002}).

\bibitem[{\citenamefont{Sides and Fredrickson}(2003)}]{sides_fred3}
\bibinfo{author}{\bibfnamefont{S.~W.} \bibnamefont{Sides}} \bibnamefont{and}
  \bibinfo{author}{\bibfnamefont{G.~H.} \bibnamefont{Fredrickson}},
  \bibinfo{journal}{Polymer} \textbf{\bibinfo{volume}{44}},
  \bibinfo{pages}{5859} (\bibinfo{year}{2003}).

\bibitem[{\citenamefont{Cochran et~al.}(2006)\citenamefont{Cochran,
  Garcia-Cervera, and Fredrickson}}]{glenn_gyr}
\bibinfo{author}{\bibfnamefont{E.~W.} \bibnamefont{Cochran}},
  \bibinfo{author}{\bibfnamefont{C.~J.} \bibnamefont{Garcia-Cervera}},
  \bibnamefont{and} \bibinfo{author}{\bibfnamefont{G.~H.}
  \bibnamefont{Fredrickson}}, \bibinfo{journal}{Macromolecules}
  \textbf{\bibinfo{volume}{39}}, \bibinfo{pages}{2449} (\bibinfo{year}{2006}).

\bibitem[{\citenamefont{Fredrickson}(2006)}]{glennbook}
\bibinfo{author}{\bibfnamefont{G.~H.} \bibnamefont{Fredrickson}},
  \emph{\bibinfo{title}{The Equilibrium Theory of Inhomogeneous Polymers}}
  (\bibinfo{publisher}{Clarendon Press, Oxford}, \bibinfo{year}{2006}).

\bibitem[{\citenamefont{Shi}(1999)}]{shi_jph}
\bibinfo{author}{\bibfnamefont{A.-C.} \bibnamefont{Shi}}, \bibinfo{journal}{J.
  Phys: Condensed Matter} \textbf{\bibinfo{volume}{11}}, \bibinfo{pages}{10183}
  (\bibinfo{year}{1999}).

\bibitem[{\citenamefont{Bouckaert et~al.}(1936)\citenamefont{Bouckaert,
  Smoluchowski, and Wigner}}]{bsw}
\bibinfo{author}{\bibfnamefont{L.~P.} \bibnamefont{Bouckaert}},
  \bibinfo{author}{\bibfnamefont{R.}~\bibnamefont{Smoluchowski}},
  \bibnamefont{and} \bibinfo{author}{\bibfnamefont{E.}~\bibnamefont{Wigner}},
  \bibinfo{journal}{Physical Review} \textbf{\bibinfo{volume}{50}},
  \bibinfo{pages}{58} (\bibinfo{year}{1936}).

\bibitem[{\citenamefont{Tinkham}(1964)}]{tinkham_book}
\bibinfo{author}{\bibfnamefont{M.}~\bibnamefont{Tinkham}},
  \emph{\bibinfo{title}{Group Theory and Quantum Mechanics}}
  (\bibinfo{publisher}{McGraw-Hill Book Company}, \bibinfo{year}{1964}).

\bibitem[{\citenamefont{Sakurai et~al.}(1993)\citenamefont{Sakurai, Kawada,
  Hashimoto, and Fetters}}]{cst_fetters}
\bibinfo{author}{\bibfnamefont{S.}~\bibnamefont{Sakurai}},
  \bibinfo{author}{\bibfnamefont{H.}~\bibnamefont{Kawada}},
  \bibinfo{author}{\bibfnamefont{T.}~\bibnamefont{Hashimoto}},
  \bibnamefont{and} \bibinfo{author}{\bibfnamefont{L.~J.}
  \bibnamefont{Fetters}}, \bibinfo{journal}{Macromolecules}
  \textbf{\bibinfo{volume}{26}}, \bibinfo{pages}{5796} (\bibinfo{year}{1993}).

\bibitem[{\citenamefont{Jones}(1975)}]{jones_book}
\bibinfo{author}{\bibfnamefont{H.}~\bibnamefont{Jones}},
  \emph{\bibinfo{title}{Theory of Brillouin zones and Electronic States in
  Crystals}} (\bibinfo{publisher}{North Holland Publishing Company, Amsterdam},
  \bibinfo{year}{1975}).

\bibitem[{\citenamefont{Tyler and Morse}(2005)}]{tyler_prl}
\bibinfo{author}{\bibfnamefont{C.~A.} \bibnamefont{Tyler}} \bibnamefont{and}
  \bibinfo{author}{\bibfnamefont{D.~C.} \bibnamefont{Morse}},
  \bibinfo{journal}{Physical Review Letters} \textbf{\bibinfo{volume}{94}},
  \bibinfo{pages}{208302} (\bibinfo{year}{2005}).

\bibitem[{\citenamefont{Tyler et~al.}(2007)\citenamefont{Tyler, Qin, Bates, and
  Morse}}]{tyler_macro}
\bibinfo{author}{\bibfnamefont{C.~A.} \bibnamefont{Tyler}},
  \bibinfo{author}{\bibfnamefont{J.}~\bibnamefont{Qin}},
  \bibinfo{author}{\bibfnamefont{F.}~\bibnamefont{Bates}}, \bibnamefont{and}
  \bibinfo{author}{\bibfnamefont{D.~C.} \bibnamefont{Morse}},
  \bibinfo{journal}{Macromolecules} \textbf{\bibinfo{volume}{40}},
  \bibinfo{pages}{4654} (\bibinfo{year}{2007}).

\bibitem[{\citenamefont{Ranjan and Morse}(2006)}]{ranjan}
\bibinfo{author}{\bibfnamefont{A.}~\bibnamefont{Ranjan}} \bibnamefont{and}
  \bibinfo{author}{\bibfnamefont{D.~C.} \bibnamefont{Morse}},
  \bibinfo{journal}{Physical Review E} \textbf{\bibinfo{volume}{74}},
  \bibinfo{pages}{011803} (\bibinfo{year}{2006}).

\bibitem[{\citenamefont{Matsen}(1998{\natexlab{b}})}]{matsen_reply}
\bibinfo{author}{\bibfnamefont{M.~W.} \bibnamefont{Matsen}},
  \bibinfo{journal}{Physical Review Letters} \textbf{\bibinfo{volume}{80}},
  \bibinfo{pages}{201} (\bibinfo{year}{1998}{\natexlab{b}}).

\bibitem[{\citenamefont{Deuflhard and Mornemann}(2002)}]{odebook}
\bibinfo{author}{\bibfnamefont{P.}~\bibnamefont{Deuflhard}} \bibnamefont{and}
  \bibinfo{author}{\bibfnamefont{G.}~\bibnamefont{Mornemann}},
  \emph{\bibinfo{title}{Scientific Computing with Ordinary Differential
  Equations}} (\bibinfo{publisher}{Springer-Verlag New York Inc.},
  \bibinfo{year}{2002}).

\end{thebibliography}
\end{document}